\documentclass[useAMS,usenatbib,a4paper,fleqn]{mn2e}

\usepackage{amsfonts,amssymb,amsmath}
\usepackage{aas_macros}
\usepackage{times,varioref,multirow,textcomp,comment}
\usepackage{xspace,float}
\usepackage[usenames,dvipsnames,svgnames,hyperref]{xcolor}
\usepackage[pdftex]{graphicx}
\usepackage[percent]{overpic}
\usepackage{}
\usepackage{comment}
\usepackage[
    pdftex,
    a4paper=false,
    plainpages=true,
    pdfpagelabels,
    breaklinks=true,
    bookmarks=true,
    bookmarksopen=false,
    bookmarksopenlevel=2
    bookmarksnumbered=true,
    bookmarkstype=toc,
    colorlinks=true,
    citecolor=RoyalBlue,
    linkcolor=ForestGreen,
    menucolor=Teal,
    urlcolor=DarkOrange,
]{hyperref}

\providecommand{\adsurl}[1]{\href{#1}{ADS}}
\def \Msun{\ {\rm M_\odot}}
\def \Msunh{\ h^{-1}{\rm M_\odot}}
\def \Mpc{{\rm Mpc}}
\def \kpc{{\rm kpc}}
\def \Mpch{\ h^{-1}{\rm Mpc}}
\def \cMpch{\ h^{-1}{\rm cMpc}}

\def \ckpch{\ h^{-1}{\rm ckpc}}

\def \LCDM{$\Lambda$CDM}

\def \vmax{V_{\rm max}}

\newcommand{\Eqref}[1]{Eq.~(\ref{#1})}
\newcommand{\Figref}[1]{Fig.~\ref{#1}}
\newcommand{\Secref}[1]{\S\ref{#1}}  
\newcommand{\Tableref}[1]{Table~\ref{#1}}

\defcitealias{nfw}{NFW}

\defcitealias{elahi2011}{{\sc stf}}

\def \velociraptor{{\sc veloci}raptor}
\defcitealias{ahf}{{\sc ahf}}

\def \Gadget2{{\sc gadget-2}}

\def\highressim{L210N1536}

\hypersetup{
  pdfauthor = {Pascal Elahi},
  pdfkeywords = {SURFS},
  pdftitle = {SURFS}
}

\def\surfs{{\sc surfs}}

\begin{document}
\title[SURFS]{SURFS: Riding the waves with Synthetic UniveRses For Surveys}
\author[P.J.~Elahi, {\it et al}.]{
\parbox{\textwidth}{
Pascal J. Elahi\thanks{E-mail: pascal.elahi@icrar.org}$^{1,2}$,
Charlotte Welker$^{1}$,
Chris Power$^{1}$,
Claudia del P. Lagos$^{1}$,
Aaron S.~G. Robotham$^{1}$,
Rodrigo Ca\~nas$^{1}$,
Rhys Poulton$^{1}$
}\vspace{0.4cm}\\ 
\parbox{\textwidth}{
$^{1}$International Centre for Radio Astronomy Research, University of Western Australia, 35 Stirling Highway, Crawley, WA 6009, Australia\\
$^{2}$ARC Centre of Excellence for All Sky Astrophysics in 3 Dimensions (ASTRO 3D)\\
}
}
\maketitle

\pdfbookmark[1]{Abstract}{sec:abstract}
\begin{abstract}
We present the Synthetic UniveRses For Surveys ({\sc surfs}) simulations, a set of N-body/Hydro simulations of the concordance $\Lambda$ Cold Dark Matter (\LCDM) cosmology. These simulations use Planck cosmology, contain up to 10 billion particles and sample scales \& halo masses down to $1~\kpc$ \& $10^8\Msun$. We identify and track haloes from $z=24$ to today using a state-of-the-art 6D halo finder and merger tree builder. We demonstrate that certain properties of haloes merger trees are numerically converged for haloes composed of $\gtrsim100$ particles. Haloes smoothly grow in mass, $\vmax$, with the mass history characterised by $\log M(a)\propto\exp\left[-(a/\beta)^\alpha\right]$ where $a$ is the scale factor, $\alpha(M)\approx0.8$ \& $\beta(M)\approx0.024$, with these parameters decreasing with decreasing halo mass. Subhaloes follow power-law cumulative mass and velocity functions, i.e. $n(>f)\propto f^{-\alpha}$ with $\alpha_{M}=0.83\pm0.01$ and $\alpha_{\vmax}=2.13\pm0.03$ for mass \& velocity respectively, independent of redshift, as seen in previous studies. The halo-to-halo scatter in amplitude is $0.9$~dex. The number of subhaloes in a halo weakly correlates with a halo's concentration $c$ \& spin $\lambda$:haloes of high $c$ \& low $\lambda$ have $60\%$ more subhaloes than similar mass haloes of low $c$ \& high $\lambda$. High cadence tracking shows subhaloes are dynamic residents, with $25\%$ leaving their host halo momentarily, becoming a backsplash subhalo, and another $20\%$ changing hosts entirely, in agreement with previous studies. In general, subhaloes have elliptical orbits, $e\approx0.6$, with periods of $2.3^{+2.1}_{-1.7}$~Gyrs. Subhaloes lose most of their mass at pericentric passage with mass loss rates of $\sim40\%$~Gyr$^{-1}$. These catalogues will be made publicly available.
\end{abstract}
\maketitle

\section{Introduction}\label{sec:intro}
Ongoing and upcoming galaxy surveys, such as {\sc alfalfa} \cite[][]{haynes2011a}, {\sc gama} \cite[][]{driver2011a,liske2015a}, {\sc waves} \cite[][]{wavessurvey,dejong2014a}, {\sc wallaby} \cite[][]{askap,wallaby} will probe galaxy formation and cosmic structure down to stellar galaxy masses of $10^{6}\Msun$ in haloes with maximum circular velocities of $10$~km/s, while other large volume surveys, like the Taipan survey (da Cunha et al., in preparation), will focus on sampling millions of galaxies to examine our cosmology. These large observational projects will sample the galaxy population with great statistics, enough to severely test our galaxy formation models. To match these surveys, typically simulation volumes should be of similar size as the survey volume. However, this is countered by the need to have high enough resolution to not only reliably identify dark matter haloes in cosmological simulations but to robustly follow their evolution. Some cosmological simulations have sacrificed resolution for survey volume \cite[e.g.][]{angulo2012,riebe2013,kim2015a,fosalba2015a} with the goal of populating pure N-body runs with galaxies using methods like Halo Occupation Distribution (HOD), combined with SubHalo Abundance Matching (SHAM) \cite[e.g.][]{zheng2005a,conroy2006a,hearin2013a,skibba2015a,carretero2015a,saito2016a}. These methods rely on simple mappings between halo masses and the galaxies that reside in them and lack the predictive power of more physical galaxy formation models such as Semi-Analytic Models \cite[SAMs, e.g.][]{cole2000,baugh2006,delucia2007a,monaco2007a,lee2013a,henriques2013a,croton2016a,lacey2016a} and full hydrodynamical cosmological simulations \cite[e.g.][]{dubois2014a,vogelsberger2014a,schaye2015a}. These techniques however, require high resolution and in the case of SAMs, accurate reconstruction of the evolution of haloes using high-fidelity halo catalogues coupled with high-cadence merger trees. Higher resolution N-body simulations with sufficient resolution (both in halo mass and the temporal resolution of the halo catalogue)  \cite[e.g.][]{springel2005,boylankolchin2009,klypin2011a} are the only means through which realistic mock surveys capable of developing galaxy formation theory can be produced \cite[e.g.][]{baugh2006,benson2010b}. 

\par
We present the next stage in these simulations, {\sc surfs} (Synthetic UniveRses For Surveys), which consists of a suite of primarily pure N-body simulations spanning a range of cosmological volumes to address both galaxy formation and cosmological surveys. Our simulation volume and resolution choices for our moderate size cosmological runs are primarily motivated by the upcoming {\sc waves-wide} survey \cite[][]{wavessurvey}, which aims to probe the stellar mass function down to a completeness limit of $M_*\sim10^{6}\Msun$ in a volume of $\sim850$ cMpc ($z<0.2$). The {\sc surfs} simulations will produce synthetic analogues of this survey, resolving dark matter haloes down to $10^{9}\Msun$, with future simulations resolving haloes down to ${10}^6\Msun$, allowing the simulation to probe stellar masses down to $\sim10^{4}\Msun$\footnote{This stellar mass limit is simply based on the stellar mass to halo mass relation roughly extrapolated to low mass galaxies \cite[e.g.][]{moster2010a,behroozi2010a,vanuitert2016a}.}. These simulations are used to produce high quality halo catalogues and halo merger trees ideal for SAMs, following in the footsteps of the Millennium simulations \cite[][]{springel2005,boylankolchin2009} and the MultiDark/Bolshoi series \cite[e.g.][]{klypin2011a,riebe2013}. Our large cosmological volume runs sacrifice resolution for larger Gpc volumes and are ideal for populating HODs/SHAM models calibrated using results from SAMs, and/or available observations \cite[see for instance][]{howlett2017b}.

\par
Here in the first of a series of papers, we focus on the properties of dark matter haloes and their evolution over cosmic time. Our goal is to use the precision tracking of cosmic structure evolution to study galaxy formation physics and satellite evolution with SAMs. Additionally, we will provide the community with free access to our halo catalogues and merger trees, useful for producing mock surveys by using these as input to galaxy formation models, a topic we will cover in upcoming papers along with our own mocks. This overview paper covers several topics, highlighting particular results at a variety of cosmological scales. 

\par 
We begin in \Secref{sec:methods} with an introduction to the simulations, analysis pipeline, data products available and discuss near-term upcoming simulations that will be released. We then focus on large-scales, specifically the cosmic web, the filaments of material connecting knots and surrounding voids in which haloes reside. In \Secref{sec:simvol}, we show what mass haloes are required to reconstruct the cosmic web and trace the web as defined by gas. We then present the $z=0$ halo and subhalo population in \Secref{sec:haloes}, focusing on numerical convergence and properties of the subhalo population. We demonstrate our halo catalogues show excellent convergence for haloes composed of $\gtrsim100$ particles and analyse the subhalo population. We end with the key analysis that can only be done with high fidelity halo merger trees from high-resolution, moderate volume cosmological simulations: an accurate reconstruction of cosmic growth and the dynamic lives of subhaloes. We show that haloes grow smoothly in dark matter mass and $R_{\vmax}$ until they begin to virialise, at which point they continue to grow in mass but become more concentrated and spherical. Subhaloes, despite being dynamic residents of haloes, show smooth internal evolution, gradually losing mass, mostly at pericentric passage. We end in \Secref{sec:discussion} \& \Secref{sec:conclusion} with a summary and discussion of {\sc surfs} and upcoming results.

\section{Methods}\label{sec:methods}

\subsection{Simulations}\label{sec:sims}
The \surfs\ suite consists of N-body simulations, most with volumes of $210\cMpch$ on a side, and span a range in particle number, currently up to 8.5 billion particles using a \LCDM\ Planck cosmology \cite[][]{planckcosmoparams2015}. The simulation parameters are listed in \Tableref{tab:sims}. Our simulations are split into moderate volume, high resolution simulations focused on galaxy formation for upcoming surveys like {\sc waves} and {\sc wallaby}, and larger volume simulations designed for surveys focused on cosmological parameters like the Taipan survey. Our moderate volume simulation parameters allows us to resolve the host haloes of galaxies with stellar masses of $10^{7.5}\Msun$ at $z=0$ with the nominal requirement that the host dark matter haloes of such galaxies be resolved with 100 particles, necessary if we are to build merger trees for coupling to Semi-Analytical Models (SAM) of galaxy formation. Our larger volume simulations parameters allow us to produce several mock surveys using halo catalogues combined with Halo Occupation Distribution (HOD) models. All simulations were run with a memory lean version of the {\sc gadget2} code on the Magnus supercomputer at the Pawsey Supercomputing Centre. 
\begin{table*}
\setlength\tabcolsep{2pt}
\centering\footnotesize
\caption{Simulation parameters}
\begin{tabular}{@{\extracolsep{\fill}}l|cccc|p{0.45\textwidth}}
\hline
\hline
    Name & Box size & Number of & Particle Mass & Softening Length & Comments\\
    & $L_{\rm box}$ [$\Mpch$] & Particles $N_p$ & $m_p$ [$\Msunh$] & $\epsilon$ [$\ckpch$] &\\
\hline
    L40N512     & $40$  & $512^3$   & $4.13\times10^7$ & 2.6 & 
                Small volume, high resolution test\\ 
    L210N512    & $210$ & $512^3$   & $5.97\times10^9$ & 13.7 & 
                Moderate volume, low resolution test\\ 
    L210N1024   & $210$ & $1024^3$  & $7.47\times10^8$ & 6.8 & 
                Moderate volume, moderate resolution\\ 
    L210N1024NR & $210$ & $2\times1024^3$ & $6.29\times10^8$ & 6.8 & 
                \multirow{2}{0.45\textwidth}{Nonradiative (adiabatic gas, no star formation or feedback) analogue to L210N1024.} \\ 
                &       &           & $1.17\times10^8$ & & \\ 
    L210N1536   & $210$ & $1536^3$  & $2.21\times10^8$ & 4.5 & 
                Moderate volume, current high resolution. \\ 
    L900N2048   & $900$ & $2048^3$  & $7.35\times10^9$ & 14.6 & 
                Large volume, low resolution, low cadence for HODs \\ 
\end{tabular}
\label{tab:sims}
\end{table*}

\par 
These simulations provide an excellent test-bed for numerical convergence, studies into the growth of haloes and the evolution of subhaloes down to dark matter halo masses of $\sim10^{10}\Msun$ (and galaxy stellar masses down to $\sim10^{8}\Msun$). Our paper here will primarily focus on the $210\Mpch$ volume (L210) simulations, though we note that our large cosmological volume simulation has already been used to produce mocks to study cosmological parameters \cite[][]{howlett2017b}. We have also run a non-radiative hydrodynamical counterpart to our L210N1024 simulation to examine the effects of gas physics and, more importantly, the rate of cosmic gas accretion, which is an essential piece of information for any SAM \cite[see review by][]{benson2010b}, although from this point we will focus on our DM only simulations. 

\par 
We produce 200 snapshots and associated halo catalogues in evenly spaced logarithmic intervals in the growth factor starting at $z=24$ for our L210 and smaller volume simulations. This high cadence, higher than was used in the Millennium simulations \cite[][]{springel2005,boylankolchin2009}, is necessary for halo merger trees that accurately capture the evolution of dark matter haloes as each snapshot is separated by less than the freefall time of overdensities of $200\rho_{\rm crit}$, i.e., haloes. 

\par 
The set of current simulations will expand to include multiple 8.5 billion particle Gpc scale simulations, along with 64 billion particle $210\Mpch$ simulations. The moderate volume, high resolution simulation will probe haloes down to mass of $3.4\times10^8\Msun$, robustly follows the cosmic evolution of $\gtrsim1.7\times10^{9}\Msun$ haloes. We will produce numerically converged synthetic galaxies down to stellar masses of $\sim10^{7}\Msun$, near the completeness limit of the {\sc waves} survey. The end goal is a 500 billion particle $210\Mpch$ simulation capable of resolving the lives of dwarf galaxies of $\sim10^{6}\Msun$. 

\subsection{Halo Catalogues}\label{sec:analysis}
We identify haloes and calculate their properties using {\sc VELOCIraptor} \cite[aka STructure Finder, {\sc stf}][Elahi et al., in prep]{elahi2011}\footnote{freely available \href{https://github.com/pelahi/VELOCIraptor-STF.git}{\url{https://github.com/pelahi/VELOCIraptor-STF.git}}}. This code first identifies haloes using a 3DFOF algorithm\footnote{We also apply a 6DFOF to each candidate FOF halo using the velocity dispersion of the candidate object to clean the halo catalogue of objects spuriously linked by artificial particle bridges, useful for disentangling early stage mergers.} \cite[3D Friends-of-Friends in configuration space, see][]{fof} and then identifies substructures using a phase-space FOF algorithm on particles that appear to be dynamically distinct from the mean halo background, i.e. particles which have a local velocity distribution that differs significantly from the mean, i.e. smooth background halo. Since this approach is capable of not only finding subhaloes, but also tidal streams surrounding subhaloes as well as tidal streams from completely disrupted subhaloes \cite[][]{elahi2013a}, for this analysis, we also ensure that a group is roughly self-bound, allowing particles to have potential energy to kinetic energy ratios of at least $0.95$. 

\par 
Like other phase-space finders, such as {\sc rockstar} \cite[][]{rockstar}, this code is better able to disentangle major mergers than configuration-space based finders \cite[][]{behroozi2015a} (like {\sc subfind}, \citealp{subfind}; or {\sc ahf},  \citealp{ahf}; see \citealp{muldrew2011} for examples of the short-comings of configuration-space halo finders). Specifically, once all deviations from the large-scale, smooth velocity distribution have been identified, a.k.a, subhaloes, the code then searches the remaining background for the cores of merger remnants, i.e., phase-space dense groups, using the velocity dispersion of the halo to scale the velocity linking lengths. If multiple cores are found, that is the smooth background is characterised by multiple large-scale phase-space distributions, then phase-space dispersion tensors are calculated. Particles in the halo background are then assigned to the closest core in phase-space as calculated using the core's phase-space tensor from the core's centre-of-mass in phase-space. This method is similar to assigning particles based on a Gaussian mixture model, but less time-consuming\footnote{The mass reconstruction using phase-space tensors for these major merger remnants can be noisy once an object becomes significantly disrupted and its particle distribution is not well characterised by a multi-variate Gaussian with a single global dispersion tensor. This flaw would also be present in full Gaussian mixture models, possibly to a greater extent, and would require generalised distribution functions to assign particles and an evaluation of the number of connections to other particles in the group.}. For a more thorough discussion of (sub)halo finding, we refer readers to \cite{onions2012,knebe2013a} and an upcoming paper on revisions to {\sc VELOCIraptor} (Elahi et al., in prep, Ca\~nas et al., in prep). 

\par
The next step is the construction of a halo merger tree. We use the halo merger tree code that is part of the {\sc VELOCIraptor} package \cite[see][, Elahi et al., in prep, for more details]{srisawat2013} called {\sc TreeFrog}. At the simplest level, this code is a particle correlator and relies on particle IDs being continuous across time (or halo catalogues). The cross-matching between catalogue $A$ \& $B$ is done by identifying for each object in catalogue A, the object in catalogue B that maximises the merit function:
\begin{equation}
    \mathcal{N}_{A_{i}B_{j}} = N_{A_{i}\bigcap B_{j}}^2/(N_{A_{i}}N_{B_{j}}),\label{eqn:merit1}
\end{equation}
where $N_{A_{i}\bigcap B_{j}}$ is the number of particles shared between objects $i$ and $j$ and $N_{A_{i}}$ and $N_{B_{j}}$ are the total number of particles in the corresponding object in catalogues $A$ and $B$, respectively. This merit function maximises the fraction of shared particles in both objects and is generally robust identifying candidate matches. However, there are instances where several possible candidates are identified. This can happen when several similar mass haloes merge at once, as loosely bound particles can be readily exchanged between haloes. 

\par 
To alleviate these issues, we follow \cite{poole2017a} and use the rank of particles as ordered by their binding energy using 
\begin{equation}
    \mathcal{S}_{A_{i}B_{j},A_{i}} = \sum_{l}^{N_{A_{i}\bigcap B_{j}}} 1/\mathcal{R}_{l,A_{i}}\label{eqn:ranking}
\end{equation}
Here the sum is over all shared particles and $\mathcal{R}_{l,A_{i}}$ is the rank of particle $l$ in halo $A_{i}$, with the most bound particle in the halo having $\mathcal{R}=1$. The maximum value of this sum when all particles are shared is $\mathcal{S}^{\rm max}_{A_{i}B_{j},A_{i}}=\gamma+\ln N_{A_{i}}$, with $\gamma=0.5772156649$ being the Euler-Mascheroni constant. 

\par
We combine \Eqref{eqn:merit1} with the normalised version of \Eqref{eqn:ranking}, i.e $\tilde{\mathcal{S}}_{A_{i}B_{j},A_{i}}=\mathcal{S}_{A_{i}B_{j},A_{i}}/\mathcal{S}^{\rm max}_{A_{i}B_{j},A_{i}}$, to obtain 
\begin{equation}
    \mathcal{M}_{A_{i}B_{j}}=\mathcal{N}_{A_{i}B_{j}}\tilde{\mathcal{S}}_{A_{i}B_{j},A_{i}}\tilde{\mathcal{S}}_{A_{i}B_{j},B_{i}}, \label{eqn:merit}
\end{equation}
where we calculate the rank ordering in both haloes in question as the rank ordering can be quite different. This would be the case for a subhalo that is completely tidally disrupted in the outskirts of a larger halo. This combined merit maximises the total shared number of particles while also weighting the match by the number of equally well bound shared particles. 

\par 
We produce a tree following haloes forward in time, identifying the optimal links between progenitors and descendants. We rank progenitor/descendant link as primary and secondary. A primary link is one where the maximum merit for a halo amongst all it candidate descendants points to a descendant which has a maximum merit amongst all its candidate progenitors that points back to the same halo, ie: the maximum merit both forward and backward. All other connections are classified as secondary links.

\par 
In an ideal case, a halo would only have one descendant and that descendant would only have one progenitor. However, identifying primary links is complicated tidal disruption and by the halo finding processes, which can lose or join haloes. In the case where a halo has been disrupted and has merged with another halo, the primary progenitor is defined as the link with the best merit in both directions and the tidally disrupted halo is flagged as a secondary progenitor with no primary descendant. If haloes have been artificially merged at a given snapshot, it is possible that the merged halo will have several possible descendants at a later snapshot. In this case, the highest merit defines the primary descendant and all other the descendants are flagged has having no primary progenitor. 

\par 
Objects with no primary progenitor will generate missing links in the tree. This problem occurs for low mass haloes that lie near the particle number threshold used by the halo finder. With fine-scale temporal resolution, these haloes appear to pop in and out of existence, leaving temporally orphan haloes in the tree. Critically, orphan subhaloes can occur at much higher masses as these can be lost by the halo finder as they pass through the dense regions of their host halo. 

\par 
This problem can be alleviated somewhat by searching multiple snapshots for candidate links \cite[see][for discussions of the pitfalls of tree-building, see for instance]{srisawat2013,behroozi2013b,avila2014a,wang2016a,poole2017a}. Here we search for primary links. If a halo does not have a primary descendant in the first snapshot, subsequent snapshots are searched till a primary link is identified or the maximum number of snapshots is reached. We typically search up to 4 snapshots, equivalent to $\approx$1~Gyr at late times and $\approx$30~Myr at early times or approximately the free-fall time at the virial overdensity, $200\rho_{\rm crit}$. For a more detailed discussion of tree building see our upcoming paper, Elahi et al., in prep. 

\par
These catalogues are freely available on request and will be made available via webserver in the near future and are ideal for input to SAMs and for following the orbits of subhaloes. In an upcoming paper, Lagos et al., in prep., we will present our mocks produced using a SAM with these catalogues as input.

\section{Simulation Volume}\label{sec:simvol}
\begin{figure*}
    \includegraphics[width=0.97\textwidth]{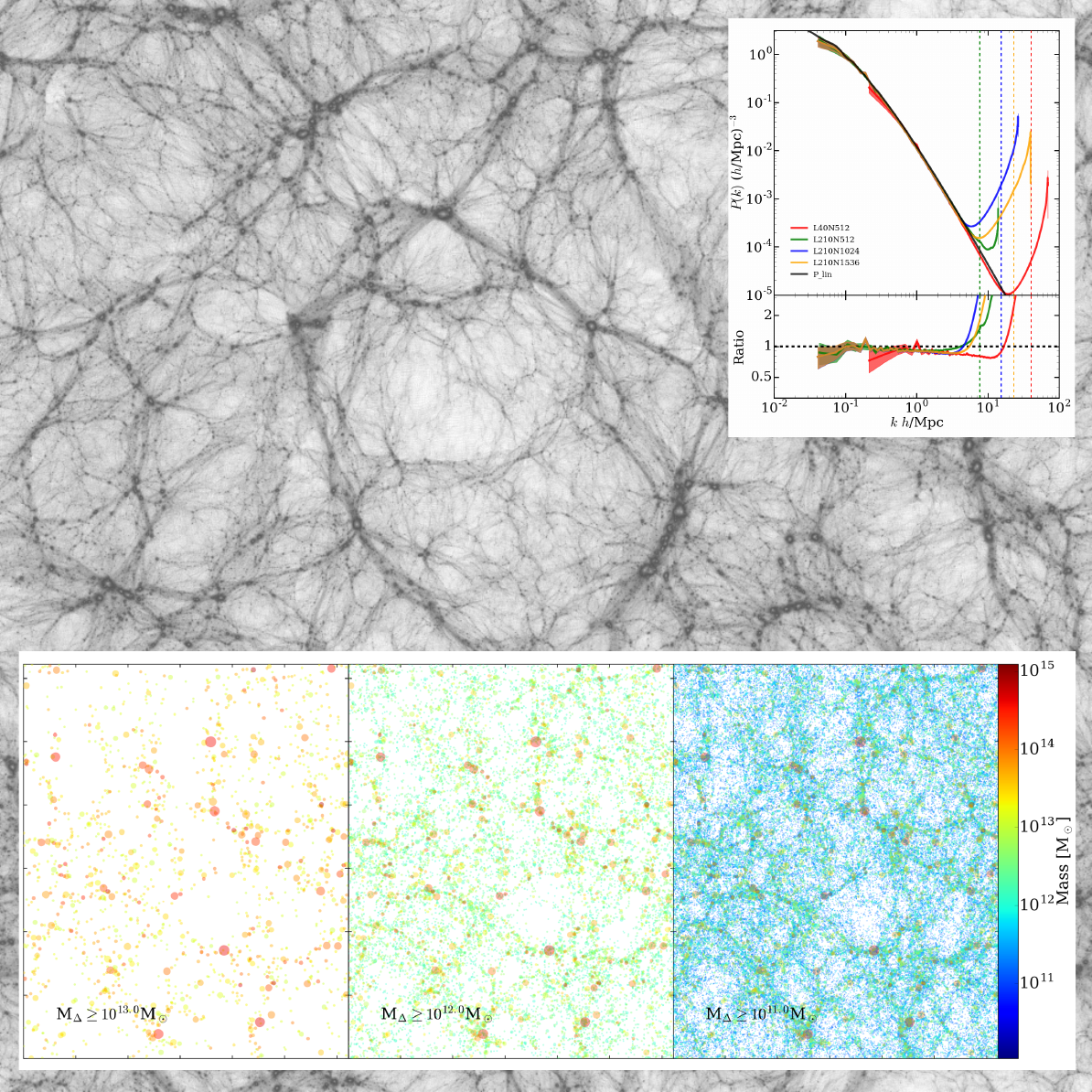}
    \caption{{\bf Simulation}. We show the density field of our highest resolution simulation, L210N1536. {\it Lower inset} shows the cosmic web as outlined by the halo distribution with different mass cuts in a slice $1/4^{th}$ the box size thick in from the L210N1536 simulation. {\it Upper inset} shows the power spectrum, $P(k)$ with $1\sigma$ sampling errors along with the ratio between all simulations and the linear power spectrum, with vertical lines at the Nyquist frequency of each simulation.}
    \label{fig:simslice}
\end{figure*}
An example of the matter distribution is presented in \Figref{fig:simslice}, where we plot the projected density field of a slice through the simulation volume from our highest resolution simulation. We also plot in the upper inset the power spectrum $P(k)=\left\langle \delta_k|\delta_k^* \right\rangle$, of our simulations at the initial conditions. A more rigorous examination of the matter power spectrum and biases produced by different tracers will be presented in later work. The take-home message here is that all our L210 volumes perfectly overlap in the initial conditions, at small $k$ values where the power spectrum is well sampled below the Nyquist frequency. The upturn in the power at early times and large $k$ is from the shot noise. 

\par 
We also plot the haloes identified by {\sc VELOCIraptor} in the lower insets, where we have applied different cuts on ``virial'' mass, here defined as $M_\Delta=4\pi R_\Delta^3 \Delta \rho_{\rm crit}/3$, with $\Delta=200$, $\rho_{\rm crit}$ is the critical density of the universe, and $R_\Delta$ is the radius that encloses this mass. Visually, we see the largest haloes appear in the densest regions of the matter field,  with smaller haloes residing in a larger variety of environments. Clearly visible in this figure and the lower insets is the cosmic web, that is a material network of nodes connected through filaments, at the intersection of walls, themselves segmenting large underdense regions, or voids. 

\subsection{Cosmic web}\label{sec:cosmic web}
The cosmic web naturally arises from the anisotropic gravitational collapse of an initially Gaussian random field of density perturbations \cite[][]{zeldovich1970,shandarin1989,peebles1980,bond1996a}. Haloes form and reside within the overdensities of the cosmic web, accreting smooth material and smaller haloes via filaments \cite[see][for details]{bond1996a}. Knots, at the intersection of several of the most contrasted filaments, house clusters; the largest virialised objects in the universe. On such scales, the filamentary pattern of the cosmic web is apparent in all large-scale galaxy surveys \cite[e.g.][]{deLapparent1986,doroshkevich2004,colless2003a,alpaslan2014a,eardley2015a}, traced by the galaxy distribution. The classification and study of this anisotropic, multi-scale cosmic density distribution is unsurprisingly a complex task. Numerous methods exist for analysing simulation data and extracting the cosmic web \cite[for an overview of various algorithmic approaches][see]{cautun2013a}. Here we do not attempt to compare different schemes nor analyse the inferred evolution of the cosmic web, leaving this for future papers. Instead we seek to answer a simple question: what haloes (galaxies) must be sampled to in order to efficiently trace the underlying gaseous structure of the cosmic web, that is the zoology of streams -cold, warm, laminar, turbulent, etc - funnelled and shaped by the cosmic web on scales that are the most relevant to galaxy formation?

\par 
Haloes, and the galaxies that reside in them, are biased spatial tracers of the cosmic velocity and density field. Studies have shown that identifying cosmic structures such as voids depends sensitively on the method and choice of tracer. Using the full, unobservable, density field will give different void regions than using galaxies with different luminosity cuts or haloes of different masses \cite[][]{paillas2016a}. The most luminous galaxies or cluster mass haloes provide information on nexus points of the cosmic web, indicating where the largest filaments terminate. Only by probing smaller halo masses or galaxies can we begin to recover more of the fine-grain features as illustrated in the inset of \Figref{fig:simslice}. Here we show the distribution of haloes for different halo mass cuts. Visually, it does not appear that major density features in this slice, like filaments, change by probing mass scales below $10^{12}\Msun$, which would correspond to haloes hosting $L^*$ galaxies. However, one should bear in mind that in such projections most walls appear as filaments and most filaments as dots, rendering any visual analysis unreliable.

\par 
To properly identify and quantify how well we recover the cosmic web, we use {\sc DisPerSe}, a topological based filament finder \cite[][]{disperse}. This algorithm identifies the ridges from a smoothed density field connecting topologically robust saddle points to peaks. {\sc DisPerSe} measures the robustness of a filament and trims the candidate catalogue in two ways: filament  persistence, the ratio of the value at the two critical points in a topologically-significant pair of critical points (maximum-saddle, saddle-saddle or saddle-minimum); and local robustness, the density contrast between the critical points and skeleton segments with respect to background. Removing low-persistence pairs is a multi-scale, non-local method to filter noise/low significance filaments. When applied to point-like distributions of haloes or galaxies, a persistence threshold translates easily into a minimal signal-to-noise ratio, expressed as a number of standard deviations $\sigma$. This algorithm has not only been used to analyse simulations \cite[e.g.][]{dubois2014a} but has been successfully applied to real spectroscopic (VIPERS) and photometric (COSMOS) surveys \cite[e.g.][]{malavasi2016a,laigle2017a,malavasi2017a}.

\par
\begin{figure}
    \centering
    \includegraphics[width=0.48\textwidth]{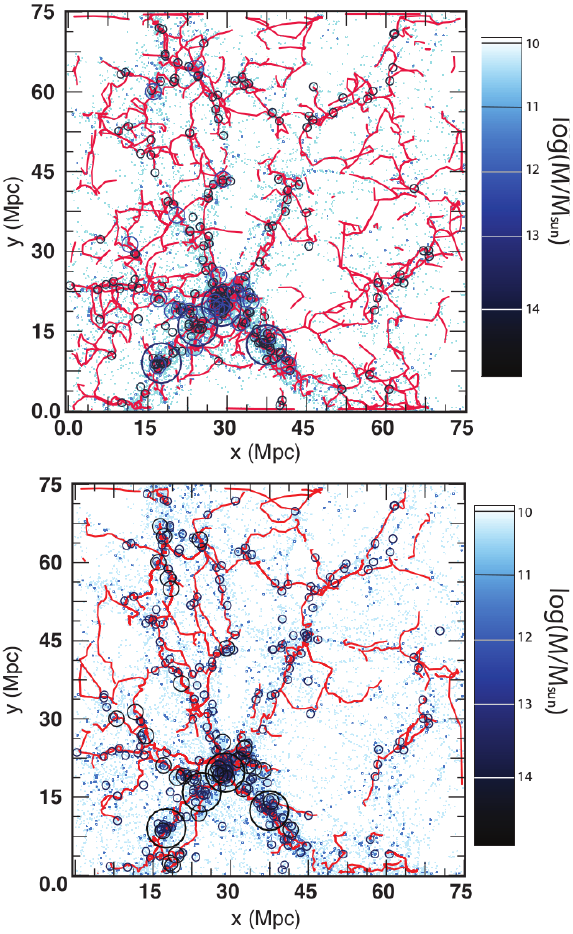}\\
    \caption{{\bf The cosmic web in SURFS:} 40 Mpc thick projected map of a subvolume of the SURFS box, with 75 Mpc on a side. The filaments extracted with {\sc DisPerSe} from the full distribution of haloes are highlighted in red while haloes in the mass range $10^{10}-10^{15}\Msun$ appear as circles of varying size and colours. {\it Top}: The persistence threshold for the cosmic web extraction is kept low, with a minimum signal-to-noise ratio of $1 \sigma$. A complicated pattern of filaments of various scales and densities is traced by the distribution of haloes. {\it Bottom}: The signal-to-noise ratio is raised to $5 \sigma$. Now only the most robust filaments appear, highlighting the smooth filamentary pattern of the cosmic web on Mpc scales.}
    \label{fig:filamentstudy}
\end{figure}

 
We apply this method on the distribution of haloes identified in different cubic sub-volumes of the simulation, with varying persistence threshold and the minimal mass of haloes considered. Let us first consider all the haloes with masses of $10^{10}-10^{15}\Msun$ identified in a $75\Mpc$ wide sub-volume of L210N1024NR. Results are presented in a 40~Mpc thick projected map in \Figref{fig:filamentstudy}. In the top panel, the persistence threshold is $ 1\sigma$ while in the bottom panel it is set to $5\sigma$. In both panels, haloes are overplotted as circles of varying colour and size depending on $\log M$. The cosmic web identified with low persistence in the left panel includes numerous short spurious filaments of widely varying directions within one thicker filament, and doubled filaments. Although in a sufficiently resolved environment, such a level of precision might be useful to resolve lower density filaments (sometimes dubbed "tendrils") in the vicinity of void galaxies, it fails to provide a simple and smooth characterisation of the large scale filaments on Mpc scale. In the bottom panel, only the most persistent filaments appear. Unlike the top panel, filaments remain coherent over a few segments in between nodes and smoothly flow to nodes, where massive haloes reside. 
\begin{figure}
    \includegraphics[width=0.48\textwidth, clip=true]{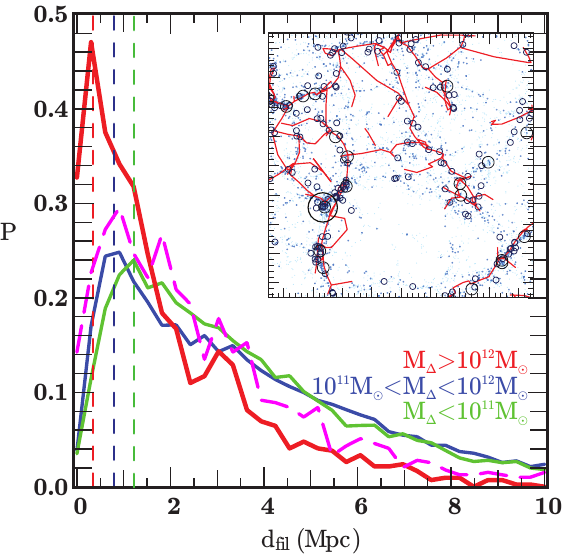}
    \caption{{\bf Halo distances to the cosmic web.} PDF of $d_{\rm fil}$, the halo distance to the nearest filament, for haloes in three different mass bins with $M_\Delta<10^{11}\Msun$ (solid green curve), $10^{11}\Msun< M_\Delta<10^{12}\Msun$ (solid blue curve) and $M_\Delta>10^{12}\Msun$ (solid red curve). Vertical dashed lines indicate the peak of the distribution. We also plot the PDF of $d_{\rm fil}$ for the sub-sample of haloes with $5\times10^{11}\Msun<M_\Delta<10^{12}\Msun$ (dashed magenta line) to examine the effect of this mass threshold on the PDF. {\it Inset} shows the skeleton extracted using haloes with with $M>5\times10^{11}\Msun$ and low persistence.}
    \label{fig:dist_fil}
\end{figure}

\par 
To determine what halo masses need to be identified in order to capture the Mpc-scale cosmic web we examine in \Figref{fig:dist_fil} the probability density function (PDF) of $d_{\rm fil}$, the distance of haloes to their nearest filament at $z=0$. Here we produce a skeleton, shown in the inset of \Figref{fig:dist_fil}, using haloes with $M_\Delta>5\times10^{11}\Msun$, keeping filaments that are significant to $\geq1\sigma$. This choice is motivated by the need to recover continuous smooth Mpc scale filaments similar to the filaments extracted from the gas density field in simulations such as Horizon-AGN \cite[][]{dubois2014a}, hence with consistent $d_{\rm fil}$ PDFs \cite[see][Chapter 3 for details]{welkerthesis}, and on  scales that can be robustly measured in observational studies \cite[][]{laigle2017a,malavasi2017a}. Such studies can identify filaments with Mpc precision, delineating voids with a typical radius of $\approx 5 - 30 \Mpc$, which is directly comparable to our reconstruction. We plot the PDF for three different halo mass bins: $M_\Delta<10^{11}\Msun$, $10^{11}\Msun<M_\Delta<10^{12}\Msun$ and $M_\Delta>10^{12}\Msun$. We also plot the PDF of $d_{\rm fil}$ for haloes with $5\times10^{11}\Msun<M_\Delta<10^{12}\Msun$ (dashed magenta curve), the mass bin right above the mass cut used to identify the filaments to quantify the effects of this threshold. This boundary bin has statistics similar to the highest mass bin: both contain roughly 2000 haloes, resulting in the peak of the PDF containing more than 100 haloes and all $d_{\rm fil}$ bins within 6 Mpc containing at least 50 haloes. They are therefore more directly comparable than the lower mass bins that contain more than 10000 haloes in total and for which bins around the peak of the PDF are resolved with more than 500 haloes. 

\par 
Haloes of $\sim10^{12}\Msun$ are strongly peaked at distances $d_{\rm fil}\lesssim2\Mpc$, with the most probable distance being $0.35\Mpc$. Haloes less massive than $10^{12}\Msun$ not only are further away, with peaks in the PDF at $d_{\rm fil}\sim1\Mpc$, twice the distance of large haloes, but have a much broader distribution. The dispersion as measured by the standard deviation is a factor of $\approx 1.5$ larger for haloes of $\lesssim10^{11}\Msun$ compared to those with masses of $\gtrsim10^{12}\Msun$. Even the bin containing haloes at the mass threshold used by {\sc DisPerSe} is not strongly peaked and is similar to the lower mass bins, except from those small deviations and extra noise due to fewer haloes per bin. This result emphasises the fact that on the scales of interest for this study (Mpc scale filaments along which galaxies drift), the cosmic web is mostly determined by those high mass haloes with $M_\Delta\gtrsim10^{12}\Msun$. Thus surveys only need to be complete down to halo masses of at least $10^{12}\Msun$, as these haloes are likely to be the nodes of the web, and ideally $5\times10^{11}\Msun$ to reliably reconstruct the large-scale cosmic web. Including small galaxies residing in low mass haloes of $<3\times10^{11}\Msun$ necessitates the use of a higher persistence (signal-to-noise) to avoid identifying tendrils. 

\par
This demonstrates the ability of surveys like {\sc waves}, which be relatively complete down to halo masses of $\sim10^{12}\Msun$, to robustly measure the cosmic web. 

\section{Halo Population}\label{sec:haloes}
We present results of our $z=0$ (sub)halo catalogues here. We start with convergence tests and then discuss the subhalo population of our simulations. 
\subsection{Halo properties \& convergence}\label{sec:convergence}
\begin{figure*}
    \centering
    \includegraphics[width=0.98\textwidth]{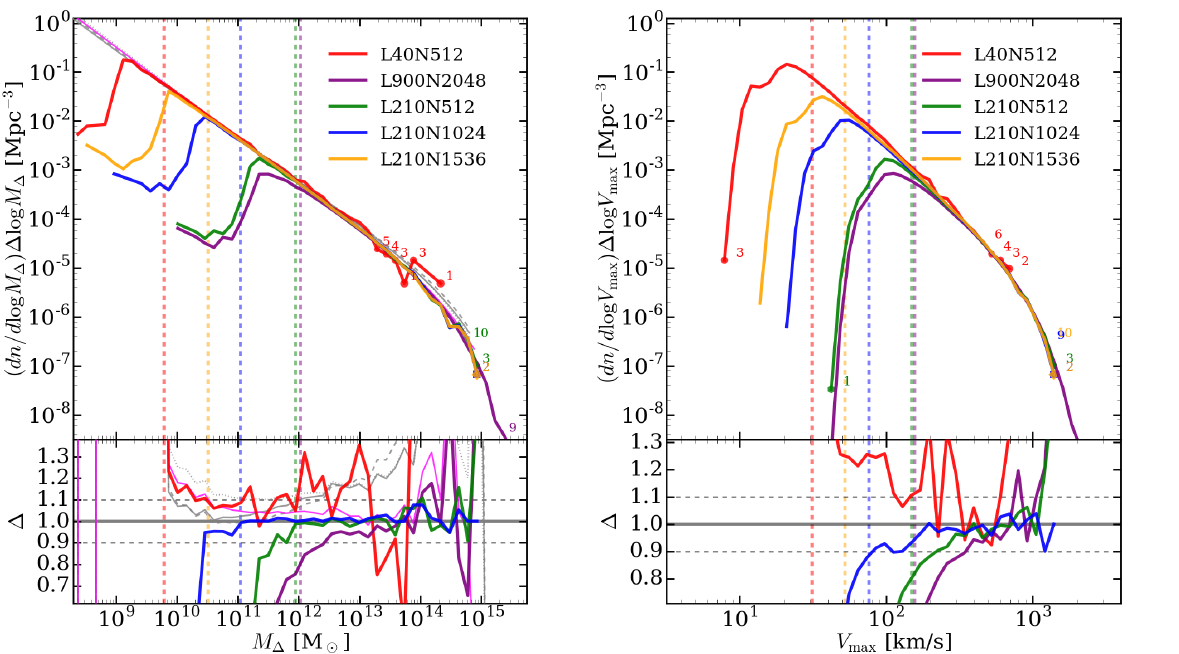}
    \caption{{\bf Halo distributions}. We show the $z=0$ halo mass and maximum velocity functions (left \& right). Each plot has the distribution in the upper panel and the residuals relative to our reference simulation in the lower panel. We highlight bins containing fewer than 10 haloes, indicating the number they contain. We also plot the mass/velocity scales of objects composed of 100 particles by vertical lines with the same colour as the corresponding simulation. We also plot four mass functions, Sheth \& Tormen (2001) (solid gray), Angulo et al., 2012 (dotted gray), the modified Angulo et al., 2012 fit keeping only bound particles in the FoF envelop, calculated using {\sc HMFCalc} (Murray et al., 2013). We also plot a fit to the highest resolution simulation (solid magenta) described in the text.}
    \label{fig:massfunc}
\end{figure*}
We start with the simplest comparison, the halo mass and velocity functions, presented in \Figref{fig:massfunc} along with the ratio of the distribution from one simulation to our highest resolution reference simulation in the bottom panels. The distribution of ``virial'' mass, here defined as $M_\Delta=4\pi R_\Delta^3 \Delta \rho_{\rm crit}/3$, with $\Delta=200$, $\rho_{\rm crit}$ is the critical density of the universe, shows that simulations with the same mass resolution give the same mass function to within $\lesssim5\%$ for haloes composed of at least 100 particles. Even with high-resolution, lower volume simulations the variance is well within $10\%$ for mass bins with Poisson fluctuations of $\lesssim20\%$. 

\par 
The sole systematic differences between simulations are a result of finite volume effects and cosmic variance. For example, larger volume simulations generally have a greater number of large haloes at mass scales above the exponential turnover, as seen by comparing the L900 simulation to the L210 simulations. Cosmic variance is easily seen in the systematic offset between L40N512 and the larger simulations. The large-scale modes present in L40N512 with wavelengths of $40\Mpch$, which are below the scale of homogeneity ($150\Mpch$ based on WiggleZ; \citealp{scrimgeour2012a}), produce an overall overdensity, enhancing halo formation.

\par 
We compare our mass functions to several fitting formulae, which all agree except at the very high mass $\gtrsim10^{13}\Msun$ where our $210\Mpch$ boxes are affected by missing power and contain fewer large haloes. We fit the binned differential mass function of our highest resolution simulation at $z=0$ using {\sc emcee} \cite[][]{emcee}. We sample the mass function at the 50 largest haloes and then 50 mass evenly spaced in $\log M$, and find parameters broadly in agreement to those found by \cite{watson2013a} (see appendix \ref{sec:halomassfunc} for more details). 

\par 
The $\vmax$ distribution (right panel of \Figref{fig:massfunc}), that is the distribution of the maximum circular velocity defined as $V^2=GM(r<R)/R$, shows similar convergence. However, the velocity scale where different resolutions diverge by $\gtrsim10\%$ occurs for haloes resolved with more than $100$ particles, unlike the mass distribution. The divergence occurs for larger haloes because internal properties like $\vmax$ require more particles before being resolved. Based on this figure, convergence of $\gtrsim95\%$ occurs for haloes composed of $\gtrsim500$ particles. 

\par
\begin{figure}
    \centering
    \includegraphics[width=0.47\textwidth,]{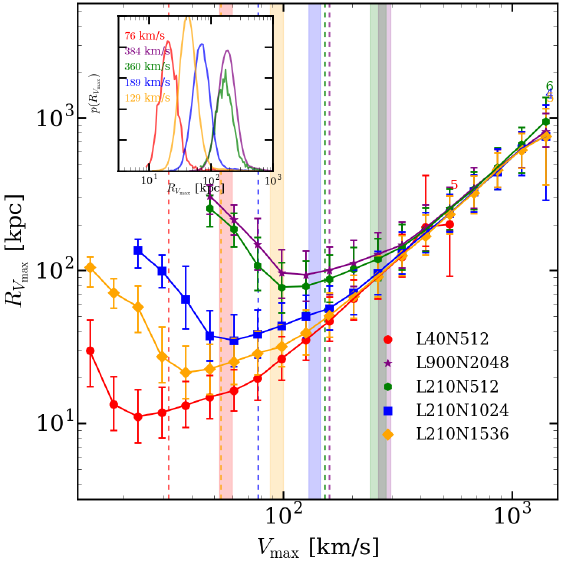}
    \caption{{\bf $\vmax$-$R_{\vmax}$ relation}. We show the median maximum circular velocity radius along with the $16\%$ \& $84\%$ quantiles in $\vmax$ bins at $z=0$. Vertical dashed lines indicate the average $\vmax$ of haloes composed of 100 particles and shaded region indicates the $2\sigma$ region of $\vmax$ values for haloes composed of 500 particles. Here we have excluded subhaloes. {\it Inset} shows the normalised distribution for haloes composed of $1000-2000$ particles for each simulation along with the median $\vmax$ within this range. Line colours are the same as in \Figref{fig:massfunc}.}
    \label{fig:vmaxrvmax}
\end{figure}

The slower $\vmax$ convergence indicates we should be cautious of the internal properties of haloes composed of $\lesssim500$ particles and is clearly demonstrated in \Figref{fig:vmaxrvmax}, where we show the radius of this region as a function of $\vmax$ for haloes only. Subhaloes are affected by strong tidal fields and hence removed from the analysis here. The simulations all show the same strong correlation between $R_{\vmax}$ \& $\vmax$ for well resolved haloes, with the median and scatter numerically converged for well resolved haloes, with the distribution in $R_{\vmax}$ at a given $\vmax$ following a Gaussian distribution as shown in the inset. Below velocity scales corresponding to $\sim500$ particles, haloes deviate away from this correlation. Thus, the internal properties of haloes composed of $\lesssim500$ particles should be treated with some caution and those composed of $\lesssim100$ particle generally ignored unless the only property one is interested in is mass. Based on this, we will typically limit our comparison between simulations to haloes composed of $\geq100$ particles, where we expect differences of at most $\sim10\%$. This limit corresponds to velocity scales of $50$~km/s in our current highest resolution L210 simulation. 

\par
\begin{figure}
    \centering
    \includegraphics[width=0.47\textwidth,]{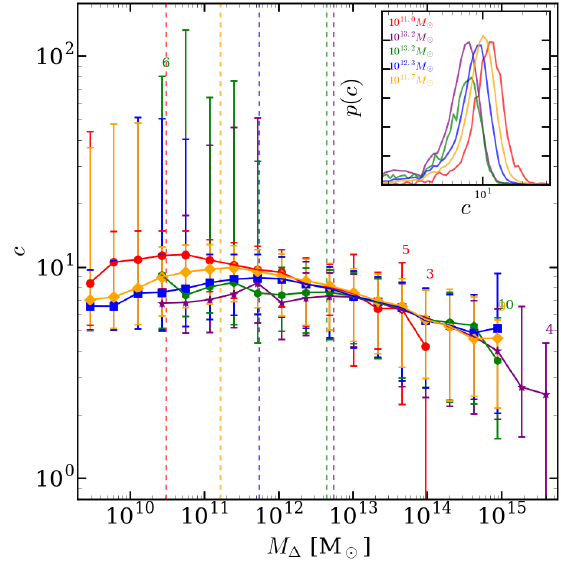}
    \caption{
    {\bf Concentration mass relation}. 
    We show the $c-M$ relation (at $z=0$) for haloes composed of $\geq100$ particles. We bin haloes in mass and determine the median and the $0.16,0.84$ quantiles for each mass bin. We also show vertical lines marking the mass scale of haloes composed of $500$ particles. {\it Inset} shows the normalised distribution of haloes composed of $1000-2000$ particles for each simulation along with the median mass within this range. Colour, marker and line styles are the same as in \Figref{fig:vmaxrvmax}.
    }
    \label{fig:concenmass}
\end{figure}

The mass profiles of dark matter haloes are reasonably well characterised by NFW or Einasto profiles \cite[e.g.][]{nfw,navarro2004}. We do not carefully fit a radial density profile using a maximum likelihood method to each halo to determine its concentration but instead follow \cite{prada2012a} and assume a NFW profile to calculate the concentration parameter $c$ via
\begin{align}
    \frac{\vmax^2}{GM_\Delta/R_\Delta}-\frac{0.216c}{\ln(1+c)-c/(1+c)}=0.
\end{align}
The concentration-mass relation of haloes is shown in \Figref{fig:concenmass}. Here we have removed subhaloes but have not removed so-called unrelaxed haloes, which are typically not well described by a NFW profile\footnote{These unrelaxed haloes are not a major issue due to the 6DFOF algorithm which removes haloes linked by particle bridges and our scheme to separate major mergers.}. We see excellent agreement between each simulation above the resolution limit of $500$ particles. The simulations reproduce the same relation, both in the mass dependence and the distribution in a given mass bin. The distribution in a given mass bin is reasonably well characterised by a lognormal distribution with a peak that increases with decreasing mass as demonstrated by the inset, where we have plotted the normalised distribution of haloes composed of $1000-2000$ particles ($0.5$~dex in mass), which corresponds to different mass scales in each simulation. 

\par
\begin{figure}
    \centering
    \includegraphics[width=0.47\textwidth,]{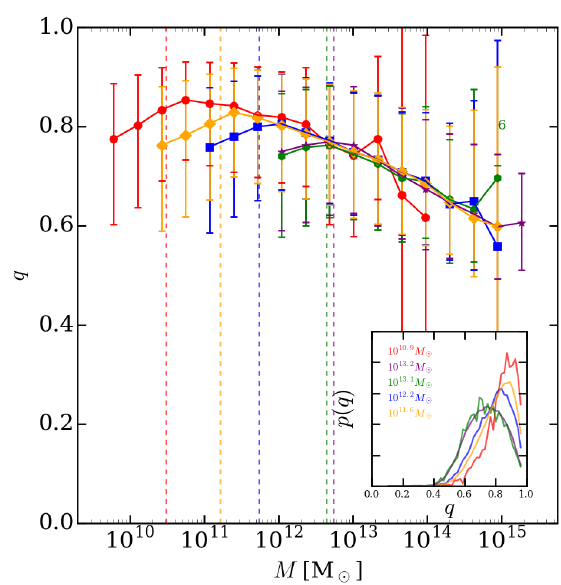}
    \caption{{\bf Shape-mass relation}. We show the semi-major axis ratio $q(M)$. Like \Figref{fig:concenmass}, we bin haloes in mass, determine medians and quantiles and show the distribution at a particular mass range in the inset. Vertical lines mark the mass scale of haloes composed of $500$ particles. Colour, marker and line styles are the same as in \Figref{fig:vmaxrvmax}.}
    \label{fig:shapeqmass}
\end{figure}

We calculate the shape using the reduced inertia tensor \cite[][]{dubinski1991,allgood2006},
\begin{equation}
    \tilde{I}_{j,k}=\sum\limits_n \frac{m_n x^\prime_{j,n} x^\prime_{k,n}}{(r^\prime_{n})^2}.
    \label{eqn:inertia}
\end{equation}
Here the sum is over particles in the halo, $(r^\prime_n)^2=(x^\prime_n)^2+(y^\prime_n/q)^2+(z^\prime_n/s)^2$ is the ellipsoidal distance between the halo's centre-of-mass and the $n$th particle, primed coordinates are in the eigenvector frame of the reduced inertia tensor, and $q$ \& $s$ are the semi-major and minor axis ratios respectively. The shape of haloes, presented in \Figref{fig:shapeqmass} shows more massive haloes tend to be more triaxial with numerical convergence occurring at the same halo resolution as that seen in \Figref{fig:concenmass}.  We only show the distribution of $q$ as the mass trend and numerical convergence for $s$ is similar, with the difference being $s\approx0.7q$. Here the addition of non-radiative gas makes large haloes more spherical by $\sim10\%$.

\par 
Overall, our simulations show excellent agreement in halo properties with strong numerical convergence for objects composed of $\gtrsim100$ particles. This is true at higher redshifts as well. 

\subsection{Subhaloes}\label{sec:subhaloes}
\begin{figure}
    \centering
    \includegraphics[width=0.47\textwidth,]{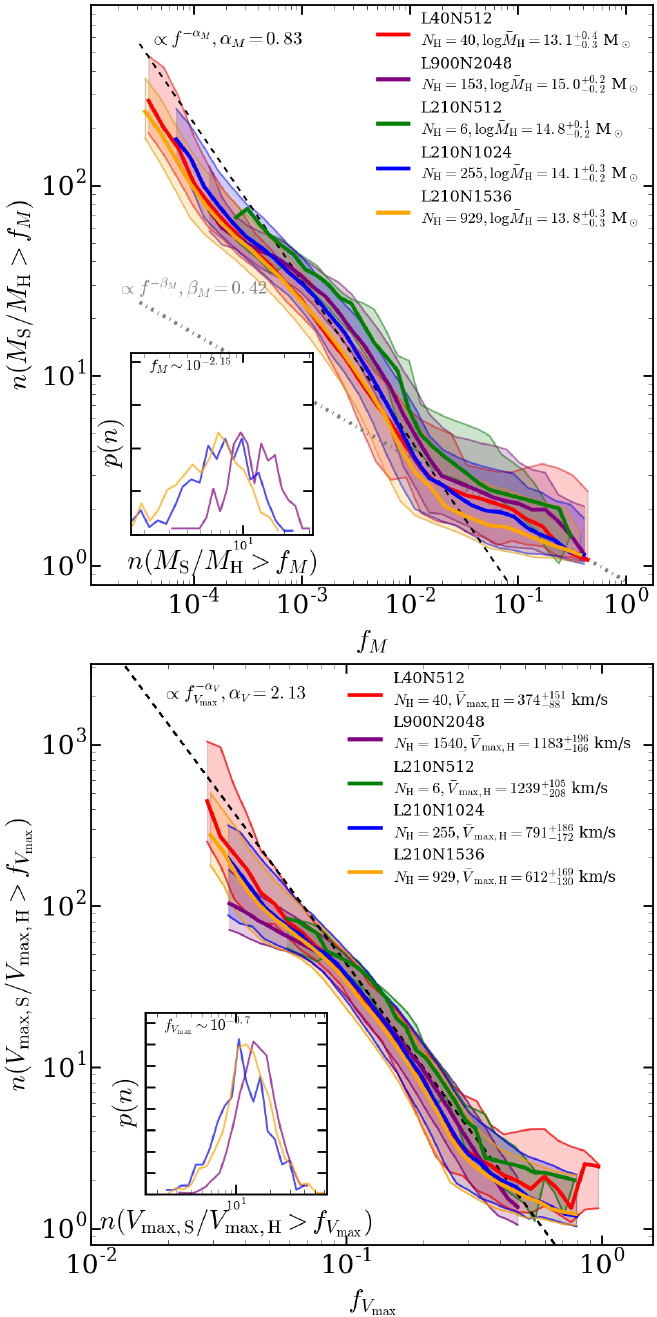}
    \caption{{\bf Subhalo distributions}. We show the $z=0$ subhalo mass and maximum velocity functions (top \& bottom). We plot the median distribution (thick line) and the $16\%$ \& $84\%$ quantiles (shaded region) using all haloes composed of more than $50000$ particles containing more than $100$ subhaloes to ensure a well sampled mass and $\vmax$ function for each halo. We also indicate the number of host haloes used and the median and $16\%$ \& $84\%$ quantiles of the host halo mass or $\vmax$. Colour, marker and line styles are the same as in \Figref{fig:massfunc}. In the insets, we show the distribution of $n$ at a given $f$ for simulations with $\geq100$ host haloes.}
    \label{fig:submass}
\end{figure}

The simulations also have numerous group mass and low cluster mass haloes that are well resolved enough to study their subhalo population. We plot the subhalo mass and circular velocity distributions in \Figref{fig:submass} for well resolved haloes containing at least 50 subhaloes and composed of 50000 particles in order to have each halo sample the subhalo mass function over a wide range of masses. Our simulations probe host masses from group scales of $\sim10^{13}\Msun$ up to small clusters of $\sim10^{14}\Msun$ (or from velocity scales of $\sim300$~km/s to $\sim800$~km/s) with our highest resolution simulation having over a thousand such haloes, with the median mass of a rich group/small cluster. To stack the distributions we normalise the mass \& $\vmax$ of subhaloes by that of their host halo, i.e., $f_M=M_{\rm S}/M_{\rm H}$ and $f_V=V_{\rm max,S}/V_{\rm max,H}$. Note here that we use the current subhalo masses and velocities, not their peak masses or $\vmax$ prior to accretion, as been sometimes done in previous work \cite[see for example][]{rodriguezpuebla2016a} and normalize by the host halo mass \& $\vmax$ excluding the contribution of subhaloes.

\par 
These plots show excellent convergence in the mass and velocity functions, with the median distribution agreeing within the scatter and, critically, the scatter is itself well converged. As has been noted many times, the distribution appears relatively scale free \cite[e.g.][]{springel2008,stadel2008}. However, unlike early studies we do not find the distribution is characterised by a simple power-law with an exponential cut-off at large masses. Instead, we see the presence of major mergers at large mass ratios of $f_M\gtrsim10^{-1}$. Haloes typically have at least one long-lived major merger remnant, with typical merger remnants having mass ratios of $0.13-0.44$ with the next largest subhalo typically a factor of $2-20$ times smaller. The relative absence of these large subhaloes in previous studies can be attributed to the use of configuration-space based finders that artificially shrink subhaloes the deeper in the parent halo they reside. The result is a mass distribution that is characterized by a double Schechter function as noted in \cite{han2017a}, which recovered these merger remnants using a tracking algorithm, HBT+.

\par 
We also see a subtle deviation in the abundance of small subhaloes with $f_M\lesssim10^{-3.25}$, which corresponds to the average mass ratio of a subhalo composed of $\lesssim100$ particles. The velocity function also shows a deviation away from a simple power-law at $f_V\lesssim9\times10^{-2}$ for all simulations resulting from poorly resolved subhaloes. These small haloes do not have well converged density profiles, $\vmax$ values, and additionally are susceptible to artificial evaporation \cite[][]{vandenbosch2017a}.

\par 
The insets show the distribution in the number of subhaloes at a given $f_M/f_{\vmax}$, i.e., the halo-to-halo scatter in the amplitude of the subhalo mass/velocity functions. The halo-to-halo scatter is roughly Gaussian. There does appear to be a shift in the most probable number of subhaloes between simulations probing different host halo mass scales. Cluster mass hosts probed in L900N2048 are richer than the group mass scales probed by our L210 simulations. 

\par
We fit the differential subhalo mass and velocity functions with a power-law functions using for all haloes simultaneously using {\sc emcee} with the log likelihood given by
\begin{equation}
    \ln L= \sum_{j}\left(-1/2 \sum_{i }\left(\frac{n_{i }- n_{{\rm model},i}}{\sigma_{i}}\right)^2-\sum_{i}\ln{\sqrt{2\pi}\sigma_{i}}\right),
\end{equation}
where $j$ is the sum over haloes, $i$ is the sum over the bins in the binned differential mass function of halo $j$, $\sigma_{i}=\sqrt{n_i+1/4}+1/2$ is the associated modified Poisson error and 
\begin{equation}
    n_{\rm model_{i}}=\int \frac{dn}{df}df=\int_{f_i}^{f_i+\delta f} \left(Af^{-\alpha}+Bf^{-\beta}\right)df,
\end{equation}
is the integral of the differential mass function over the bin. For the purposes of fitting, we do not fit the double power-law simultaneously, instead we first limit the fit to the steeper index $\alpha$, corresponding to substructures with well defined dynamical masses as opposed to objects with masses of $\gtrsim5\%$ of host halo's mass that have been flagged by \velociraptor\ as possible merger remnants. We only fit well resolved haloes with large subhalo populations and limit the fit to subhaloes composed of at least $100$ particles. We also try fitting each halo individually to asses the halo-to-halo scatter, which is much larger than the scatter on the median parameters arising from fitting all haloes at once. 

\par 
We find the average power-law for the mass function to be $\alpha_{M}=0.83\pm0.01$ for our highest resolution simulation. Fitting each halo individually gives $\alpha_{M}=0.77\pm0.26$, in agreement with the fit to the average, to other {\sc surfs} simulations, and previous estimates \cite[e.g.][]{madau2008,springel2008,stadel2008,gao2012,onions2012,rodriguezpuebla2016a,han2017a}. The amplitude from fitting each halo individually is $A_M=0.11^{+0.82}_{-0.10}$, indicating halo-to-halo scatter of $\sim0.9$~dex, although the overall scatter inferred from fitting each halo individually is likely an overestimate as each halo has few subhaloes, so the scatter in each fit is dominated by the poor constraining power of each halo. We also note that the amplitude is highly correlated with $\alpha$ (higher amplitudes, lower $\alpha$ values). 

\par 
The flatter high-mass fraction tail, corresponding to minor and major merger remnants, is less well defined because these remnants are comparatively rarer than subhalo accretion, resulting in poor sampling. Only on average does this region look like a power-law, with a poorly constrained slope of $\beta=0.42^{+0.55}_{-0.32}$ \footnote{\cite{han2017a} also found a wide range of slope values depending on the mass used, the peak or current mass, mass enclosing 200 times the critical density or 200 times the mean density.}. This region is better characterised by a (possibly skewed) Gaussian distribution in $\log f$, representing the distribution of ratios of merger events. By fitting the average current subhalo mass function, we are estimating the average mass ratios of long-lived mergers. We find $\bar{\log f}_{\rm mergers}=-0.83\pm0.01$, with a the dispersion is $\sigma_{{\log f}_{\rm mergers}}=0.312\pm0.003$, that is the typical merger remnant has a mass ratio of $7-30\%$.

\par 
For the velocity function, limiting the fit to subhaloes composed of $\geq100$ particles constrains it to $f_{\vmax}\gtrsim10^{-1}$. We find the average distribution has a steep slope of $\alpha_{V}=2.13\pm0.03$, while the fit to each halo independently gives $\alpha_{V}=2.09\pm0.86$. 

\par 
A cursory glance of the subhalo distribution at various redshifts suggests that the indices show little evolution with redshift. We leave a more robust statistical analysis of the intrinsic scatter and evolution for future papers, improving subhalo distribution models \cite[such as presented in][]{han2016a}.

\par
\begin{figure}
    \centering
    \includegraphics[width=0.49\textwidth,]{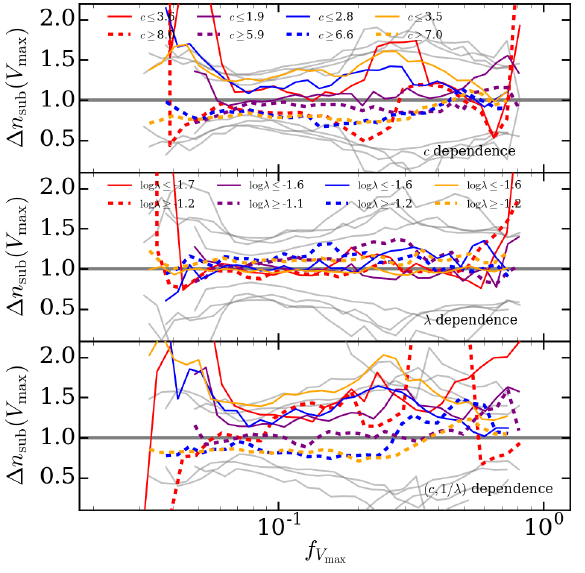}
    \caption{{\bf Subhalo distribution dependence}. We show the $z=0$ subhalo maximum velocity function ratio split according concentration $c$, spin $\lambda$ and both (top, middle \& bottom respectively). In all panels we compare the subhalo distribution in the lower (solid thin line) and upper (dashed thick line) $25\%$ to the median distribution within $25\%-75\%$. Gray lines indicate the scatter relative to the mean within the $25\%-75\%$ quantile region. Colours are the same as in \Figref{fig:massfunc}.}
    \label{fig:subvmaxdep}
\end{figure}

The amount of substructure shows significant halo-to-halo scatter but does it correlate with other bulk halo quantities? Does the subhalo abundance show some form of assembly bias? We find a correlation between the amount of substructure and halo's  concentration and spin. This dependence is presented in \Figref{fig:subvmaxdep}, where we split the host halo population by $c$ \& $\lambda$ and compare the subhalo $\vmax$ distribution of haloes between haloes split according to these two properties. Specifically we compare haloes in lower (upper) $25^{\rm th}$ percentiles to the median distribution of haloes within these percentiles, i.e., average haloes, plotting the ratio of the number of subhaloes within a given $\vmax$ range. We also show the scatter in the number of subhaloes within this median bin, indicative of halo-to-halo scatter.

\par 
The upper panel of \Figref{fig:subvmaxdep} splits haloes according to concentration. This panel shows concentrated haloes have more subhaloes than less concentrated ones across all $\vmax$ scales (solid lines compared to dashed lines), having typically $\sim30\%$ more subhaloes. However, we note the halo-to-halo scatter (solid gray lines) encompasses this difference. 

\par 
The middle panel splits haloes according to spin. Here we see the reverse trend (dashed lines above solid lines): haloes with low spin have more subhaloes than those with high spin, having typically $\sim30\%$ more. Again here the systematic bias is generally within the scatter of typical haloes. 

\par
The last panel shows haloes split according to both properties. In this case, the systematic difference between concentrated, low spin haloes and diffuse, high spin haloes (bottom) panel is tentatively significant (to $\sim1\sigma$ for $f_{\vmax}\gtrsim10^{-1}$) with a difference of $\sim60\%$. A principle component analysis of concentration, spin, shape, and maximum circular velocity is inconclusive, showing the scatter in the number of subhaloes is not strongly dominated by any particular property, with some dependence on $c$ \& $\lambda$. The number of subhaloes does leave an imprint on the global halo properties but this imprint is weak at best and a group/cluster's current bulk properties are not necessarily a strong indicator of a rich or poor group/cluster environment. However, it is possible that other evolutionary quantities are strongly correlated with the present day richness, a topic we will explore in upcoming work.

\section{Cosmic Growth}\label{sec:growth}
The simple picture of cosmic mass growth is one in which haloes grow in mass via smooth mass accretion and through the tidal disruption of subhaloes, and subhaloes slowly lose mass as they are pulled towards the centre of their host via dynamical friction till they are completely disrupted. However, this neglects tidal fields produced by nearby haloes, subhaloes leaving their host halo as they near apocentre, and major mergers, which can excite particles resulting in some mass (and angular momentum) loss. We analyse the growth of haloes and the evolution of subhaloes in this section but begin with convergence tests. 

\par
We note that the last 4 snapshots have not been fully corrected as we have not evolved our simulations in the future. However, this only spans $z=0$ to $z=0.028$, a small fraction of cosmic time. Moreover, only $\approx1.5\%$ of haloes composed of $\lesssim100$ particles are affected and this percentage drops to $\lesssim0.1\%$ for haloes composed of $\gtrsim500$ particles (see following section).

\subsection{Numerical Convergence}
\begin{figure}
    \centering
    \includegraphics[width=0.47\textwidth,]{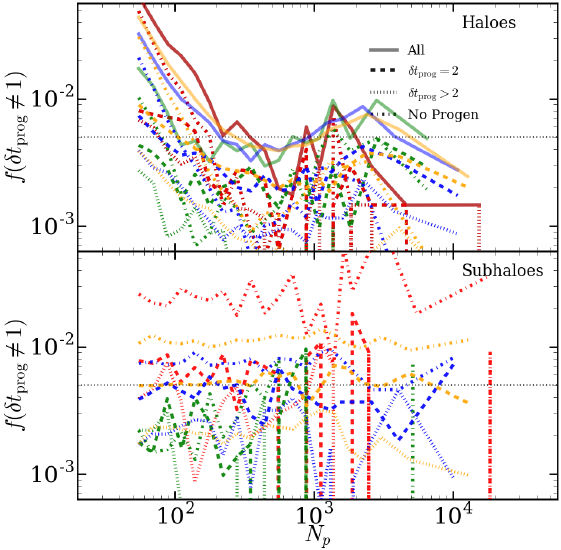}
    \caption{{\bf Halo fractions missing immediate progenitors}. We show the fraction of haloes (top) \& subhaloes (bottom) that have do not have immediate progenitors as a function of the number of particles of which a (sub)halo is composed. Specifically we show the fraction that have a progenitor identified two snapshot ago (dashed lines), more that two snapshots ago (dotted lines), and no progenitor at all (dash-dotted lines), along with the total fraction missing an immediate progenitor (solid). Colours are the same as in \Figref{fig:massfunc} and line styles are given by the legend.}
    \label{fig:progenitordeltat}
\end{figure}
\begin{figure}
    \centering
    \includegraphics[width=0.47\textwidth,]{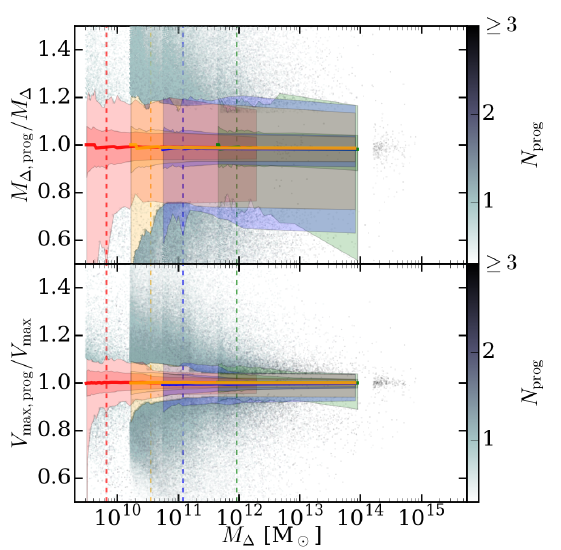}
    \caption{{\bf Progenitor properties}. We show distribution of progenitor to descendant mass \& circular velocity (top \& bottom respectively) as a function of halo mass. We plot the median ratio (thick line), the $16\%$ \& $84\%$ quantiles (shaded region), and $5\%$ \& $95\%$ quantiles (light shaded region) using all haloes composed of $\geq50$ particles. We also plot outliers and colour code these points by the number of progenitors they have. Vertical dashed lines indicate the average virial mass at which (sub)haloes are composed of 100 particles. Colour, marker and line styles are the same as in \Figref{fig:massfunc}.}
    \label{fig:progenitorratio}
\end{figure}
We examine how well we recover (sub)halo evolution using several diagnostics. We first start with comparing the properties of a (sub)halo to its immediate progenitor. The time difference between halo and progenitor is a minimum of $\sim200$~Myr, but can be up to $1$~Gyr for haloes which have an optimal progenitor found 4 snapshots in the past. This is rare and only occurs for $\sim0.3\%$ of haloes at all cosmic times. Typically only $\sim1-2\%$ of haloes have progenitors found more than a single snapshot in the past or not found at all. We show the dependence on halo particle number in \Figref{fig:progenitordeltat}. There is a strong resolution dependence on the fractions of {\em haloes} missing immediate progenitors, present in all our simulations regardless of mass resolution. There maybe a subtle halo mass dependence on the fractions themselves, with simulations probing lower mass scales having larger fractions of less than ideal progenitor links (going from L210N512 given by the green curve to L210N1536 given by the orange curve for example). These less than ideal links or missing links drops to $\lesssim1\%$ for haloes composed of $\gtrsim100$ particles. Large haloes that do not have ideal progenitors occur in multi-merging systems.

\par 
{\em Subhaloes}, shown in the lower panel, display different behaviour due to the highly nonlinear, tidally disruptive environment in which they live. Here, no resolution dependence is seen and $2\%$ of all subhaloes have less than ideal links, evenly split between finding a progenitor two snapshots in the past and not finding a ideal progenitor within four snapshots. The subhaloes with missing progenitors are typically major merger events, and occasionally multi-mergers. In upcoming work, we describe further improvements to tree building combined with additional particle tracking that necessary to fix these rare cases (Elahi et al, in prep, Poulton et al, in prep).

\par 
Next we investigate the progenitor's properties compared to the current (sub)halo in \Figref{fig:progenitorratio}, specifically the mass \& $\vmax$ ratio between a (sub)halo and its progenitor. The thick lines show the median ratio of progenitor to descendant mass \& $\vmax$ for each simulation, with the shaded regions showing the scatter. For all mass scales, most objects do not evolve significantly over this period, with the median being $\approx1$ to within $\lesssim1\%$. If we separate subhaloes from haloes we find that the median ratio is $\approx0.99$ whereas for subhaloes the median ratio is $\approx1.01$: unsurprisingly haloes grow and subhaloes shrink on average. The median and, more  importantly, {\it the scatter} in this evolution is the same for all simulations for haloes composed of $\geq500$ particles. Clearly the merger history of objects is only strongly numerically converged for objects above this particle limit. Even haloes composed of 100 particles have a the $1\sigma$ scatter $\sim10\%$ larger than better resolved haloes, showing this numerical effect. The scatter in the mass ratio is larger than the scatter in $\vmax$, $\Delta M_{\rm prog}/M\approx5\%$ compared to $\Delta V_{\rm max,prog}/V_{\rm max}\approx2\%$. The larger scatter is a result of the mass loss and growth mechanisms such as accretion of mass from subhaloes and subhaloes leaving their host having a larger impact on the outer regions of halo relative to the central regions defined by $R_{\vmax}$, despite the longer dynamical times. The asymmetry in the $2\sigma$ scatter is a result of major mergers, which can drastically increase the mass of a halo. 

\par 
Changes in the mass accretion rate between two consecutive steps are also informative. We follow \cite{contreras2017a} and define this change as 
\begin{align}
    \delta\Gamma=\frac{|\Delta_{i,i+1}M_\Delta+\Delta_{i+1,i+2}M_\Delta|}{|\Delta_{i,i+1}M_\Delta|+|\Delta_{i+1,i+2}M_\Delta|},
\end{align}
where we average over steps $i$,$i+1$, \& $i+2$. This ratio is only $<1$ in instances where the halo undergoes a mass decrease (increase) followed by an increase (decrease). \cite{contreras2017a} found it necessary to clean Millennium (Bolshoi) catalogues of those haloes with $\delta\Gamma\leq0.1$ ($\delta\Gamma\leq0.3$), haloes which experienced significant  and likely artificial fluctuations in the accretion rate. The Millennium catalogues was particularly affected across all halo masses according to \cite{contreras2017a}, even for haloes composed of 1000s of particles. 

\par 
We find only haloes above $\sim200$ particles show little mass dependence in $\delta\Gamma$, with $6\%$ of haloes with $\delta\Gamma\leq0.1$. This mass accretion rate change for haloes composed of fewer particles show a strong dependence on particle number, increasing to $12\%$ for haloes composed of fewer than 50 particles, indicating mass accretion convergence for haloes composed of $\gtrsim200$ particles. Moreover, this test indicates {\sc surfs} halo catalogue is not as affected by spurious mass accretion changes as the Millennium catalogues.

\par 
The results are the same at higher redshifts, indicating convergence in reconstruction of merger histories.

\subsection{Halo Evolution}\label{sec:haloevo}
\begin{figure}{\color{white}[I can see you]}
    \centering
    \includegraphics[width=0.49\textwidth,]{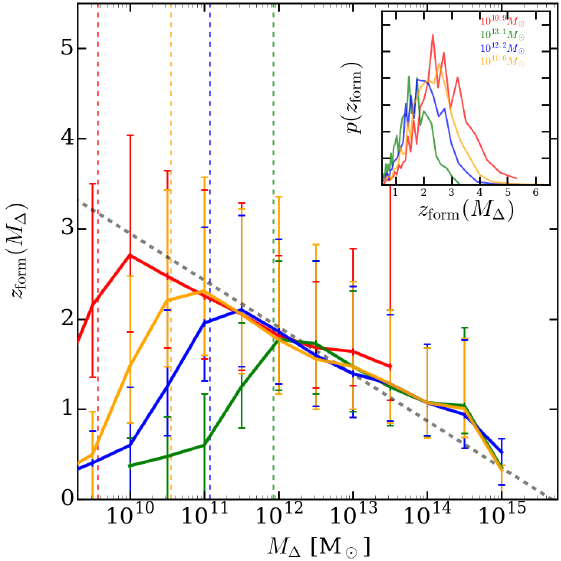}
    \caption{{\bf Halo Formation Time}. We show the halo formation time as a function of mass dependence. Like \Figref{fig:concenmass}, we bin haloes in mass, determine medians and quantiles and show the distribution at a particular mass range in the inset. Vertical lines mark the mass scale of haloes composed of $100$ particles. Colour, marker and line styles are the same as in \Figref{fig:vmaxrvmax}. We also show a fit to the distribution by a dashed gray line.}
    \label{fig:aform}
\end{figure}
\begin{figure}{\color{white}[I can see you]}
    \centering
    \includegraphics[width=0.49\textwidth,]{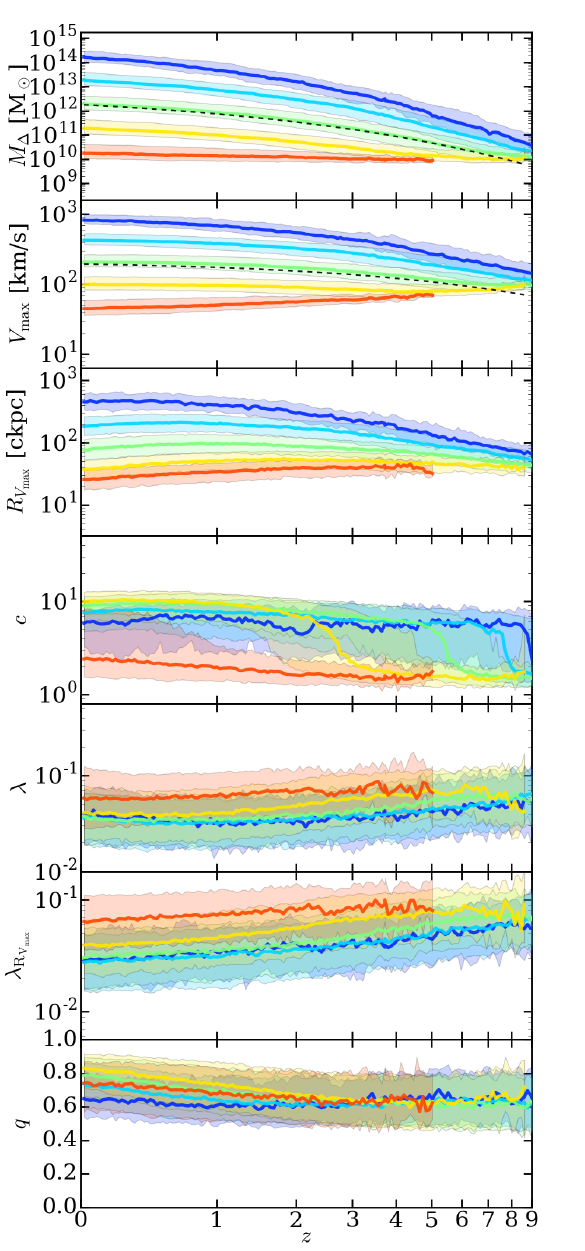}
    \caption{{\bf Halo Evolution in L210N1536}. We show the average evolution of haloes in different $z=0$ mass bins. For each bin we plot the median evolution (solid thick lines) and shaded regions showing the $16\%,84\%$ quantiles. Colours go from blue to red in decreasing $M_\Delta(z=0)$. We show the mass, $\vmax$, comoving $R_{\vmax}$ size, concentration, spin at the virial radius, the spin parameter within $R_{\vmax}$ and the shape, here just given by the semi-major axis ratio, $q$.}
    \label{fig:avehaloevo}
\end{figure}
\subsubsection{Formation Time}
We start with the formation time of haloes, which is itself a useful numerical convergence diagnostic. We show the formation time, here defined as the redshift $z_{\rm form}$ at which a halo's $M_\Delta(z)$ has $25\%$ of its current day $M_\Delta$. The formation time monotonically decreases with increasing present day halo mass. For all well resolved haloes composed of $\gtrsim200$ particles, all simulations show the same median and scatter. For haloes composed of $\lesssim100$ particles, the formation time becomes biased to lower redshifts and later times. The $z_{\rm form}-M_\Delta$ relation is well characterised by $z_{\rm form}=\alpha\log M_\Delta+\beta$, with $\alpha=-0.5\pm0.001$ \& $\beta=8.15\pm0.001$ when fitting the full halo population (not the binned data plotted here), similar to the results of \cite{power2012} although they looked at the formation time for $50\%$ of the current day mass.

\subsubsection{Dark Matter Growth}
The average evolution of {\em haloes} is presented in \Figref{fig:avehaloevo} for the \highressim, where we have removed subhaloes, which have very different evolutionary paths. Here we split haloes into $z=0$ mass bins a decade in size from $10^{9}\Msun$ to $10^{15}\Msun$ and for completeness we have included haloes down to $20$ particles at $z=0$. For each mass bin we calculate the median evolutionary track along with $1\sigma$ quantiles. 

\par 
Figure \ref{fig:avehaloevo} clearly shows smooth mass evolution with the largest haloes having accreted $10\%$ of their $z=0$ mass by $z=2$. The mass bin that differs in evolution, the red curve, corresponds to haloes composed of 20-90 particles, hence the lack of mass growth due to progenitors being below the resolution limit. If we examine the same mass bin in higher mass resolution simulations such as L40N512 where this bin contains haloes composed of 163-1634 particles, we find that the mass growth is the same as haloes of larger mass (see \Figref{fig:avehaloevoL40N512}). The $1\sigma$ scatter in the evolution is $\approx0.3$~dex. The initial high mass growth at early times followed by a turnover at late times has been noted in numerous studies \cite[e.g.][]{wechsler2002a,vandenbosch2002a,tasitsiomi2004a,mcbride2009a,rodriguezpuebla2016a}.

\par 
Several parametrisations of halo mass growth have been used, from single parameter exponential growth \cite[][]{wechsler2002a} to two parameter models \cite[][]{vandenbosch2002a,tasitsiomi2004a}. We use a three parameter model, a simpler functional form than that proposed by \cite{rodriguezpuebla2016a} to characterise the average growth of haloes of mass $M_{\Delta,o}(a=1)$, 
\begin{align}
    \log M_\Delta(a)=A(M_{\Delta,o})\exp\left[-(a/\beta(M_{\Delta,o}))^{-\alpha(M_{\Delta,o})}\right],
\end{align}
where optimal fit parameters, $A$, $\alpha$ \& $\beta$ depend on the final mass of the halo in question. Our fits are listed in \Tableref{tab:halogrowth}. In general we find, parameters decrease with decreasing halo mass, although he statistical significance of this trend is low due to the scatter in the average evolution. Additionally, the smaller mass bins have artificially flatten growth rate as smaller progenitors lie below the resolution threshold. The growth rate, $dm/da=\alpha\beta^{\alpha+1}a^{-(\alpha+1)}$ decreases with increasing $a$. 

\par
The $\vmax$ curves trace the mass growth, monotonically increasing with time save for mass bins of poorly resolved haloes. Comparing this figure to \Figref{fig:avehaloevoL40N512}, we find that only mass bins of haloes composed of $\gtrsim500$ particles show numerically converged evolution. The average scatter in evolutionary paths across cosmic time is $\approx0.1$~dex. Again, the growth can be fit by a similar function to the mass growth and the fits are also listed in \Tableref{tab:halogrowth}. The inferred growth rate of $\vmax$ is smaller than the mass growth rate.
\begin{table}
\setlength\tabcolsep{2pt}
\centering\footnotesize
\caption{Halo Growth}
\begin{tabular}{cccc}
\hline
\hline
    & $A$ & $\alpha$ & $\beta$ \\
\hline
    Mass ($\Msun$) \\
\hline
    $1.7\times10^{14}$ & $8.15_{-0.02}^{+0.07}$ & $0.80_{-0.09}^{+0.08}$ & $0.028_{-0.03}^{+0.02}$ \\
    $1.7\times10^{13}$ & $8.01\pm0.05$ & $0.73\pm0.09$ & $0.020_{-0.02}^{+0.04}$ \\
    $1.8\times10^{12}$ $^{\dagger}$ & $7.87_{-0.07}^{+0.21}$ & $0.58_{-0.23}^{+0.13}$ & $0.013_{-0.005}^{+0.21}$ \\
    $1.7\times10^{11}$ $^{\dagger\dagger}$ & $9.65_{-1.75}^{+0.11}$ & $0.062_{-0.01}^{+0.21}$ & $0.320_{-0.315}^{+0.472}$ \\
\hline
    $\vmax$ (km/s) \\
\hline
    $820$ & $4.49\pm-0.05$ & $0.95_{-0.15}^{+0.17}$ & $0.034_{-0.004}^{+0.005}$ \\
    $413$ & $4.19\pm-0.05$ & $0.91\pm0.15$ & $0.025_{-0.004}^{+0.005}$ \\
    $203$ $^{\dagger}$ & $3.96_{-0.07}^{+0.21}$ & $0.71\pm0.15$ & $0.013_{-0.006}^{+0.009}$ \\
    $97$ $^{\dagger\dagger}$ & $3.50_{-0.02}^{+0.05}$ & $1.3_{-0.7}^{+0.9}$ & $0.017_{-0.0012}^{+0.026}$ \\
\hline
\end{tabular}
\label{tab:halogrowth}
\\$^{\dagger}$Affected by resolution. $^{\dagger\dagger}$Severely affected by resolution. 
\end{table}

\par
The comoving size of a halo as defined by $R_{\vmax}$ also shows smooth evolution with little scatter. However, unlike mass and $\vmax$, the comoving size does not grow continuously but instead peaks at mass-dependent redshifts and gradually decreases over time. This is not a result of the use of comoving gravitational softening lengths, nor is the turnover resolution dependent except for poorly resolved halo as can be seen by again comparing this figure to \Figref{fig:avehaloevoL40N512} in \Secref{sec:cosmicgrowthapp}. The redshift at which this turnover occurs depends on mass with smaller mass haloes found at $z=0$ turning over at higher redshift. The turnover corresponds to when the mass variance $\sigma(M)\approx1.0$ at the average progenitor mass, that is when haloes are common nonlinear regions. As these scales become more nonlinear, the haloes tend to virialise, becoming more concentrated and the comoving size shrinks. For halo masses of $\lesssim10^{12}\Msun$, $R_{\vmax}$ decreases by $\sim50\%$ from its peak size. 

\par 
The median evolution for concentration, spin and shape are generally a simple function of $z$. However, these quantities show significant scatter in evolutionary tracks, on the order of $0.4$~dex. The abrupt change in the median concentration seen at high redshift for large masses is a result of haloes being on average resolved enough for physically meaningful concentrations using maximum circular velocities to be calculated ($\sim50$ particles). The upturn in $c$ \& $\lambda$ occurs at roughly the same time as the downturn in $R_{\vmax}$.

\par 
In general, key trends are that haloes become more concentrated at late times, with the comoving size shrinking below $z=3$ with large haloes identified today contracting later than smaller haloes. The angular momentum does not evolve significantly, although the internal angular momentum of large haloes steadily decreases with time. Current day haloes at a given mass are less triaxial and more compact that their high $z$ counterparts at similar halo mass.

\subsection{Subhalo Evolution}\label{sec:subhaloevo}
\subsubsection{Do subhaloes leave home?}
It has been known for a while that subhaloes live interesting lives. They can momentarily leave their host (so-called backsplash galaxies), exchange hosts, are subject to the tidal field of their host, and encounter other subhaloes \cite[e.g.][]{knebe2011b}. \cite{vandenbosch2017a} recently outline 12 evolutionary paths for subhaloes, most physical, others due to the limitations of the temporal resolution at which subhaloes are identified and the biases of the halo finder. The dominant evolutionary channel is one in which a subhalo continues to be a subhalo from one output to the next, followed by those that momentarily leave their host. But how often does this happen over the lifetime of a subhalo? Are subhaloes sedentary, rarely leaving a host halo once accreted? 

\par 
We examine the lives of subhaloes by following subhaloes identified at $z=0.4$ (4 Gyrs ago, to allow subhalo lifetimes to be measured) back to their accretion and then following their lives afterwards. We only examine subhaloes that were composed of $\geq100$ when first accreted and limit our analysis to host haloes that have a subhalo population of $\geq10$, that is moderately resolved host haloes and do not split results based on host halo mass nor examine any dependence on cosmic time, leaving a more detailed analysis for a future study. These selection criteria mean that different simulations explore different mass scales.

\par 
We define flybys as subhaloes that were subhaloes for at most 4 snapshots (or less than the freefall time at $R_\Delta$) over the course of their life. Haloes that swap hosts are, naturally, subhaloes that have changed host at least once since being first accreted. Here we define backsplash subhaloes as those that momentarily leave their host halo\footnote{Here we mean by ``leave a halo'' as a subhalo that is no longer in the 6DFOF envelop defining a halo. The exact definition is not critical but it is worth noting that there are several definitions of the edge of a halo and what constitutes a subhalo. The typical halo edge is commonly delineated by the virial radius, which has several definitions in the literature. For example, \cite{more2015a} discuss some of these definitions, present some of their drawbacks and advocate using the ``splashback'' radius, the mean apocentre of particles belonging to the halo as a more physical boundary to a halo. This radius is related to $R_\Delta$ and more critically, this first caustic surface is likely related to the 6DFOF envelop used to define our haloes, hence we feel justified here to use our definition of subhalo.}, excluding those subhaloes that have also swapped hosts. We define preprocessed subhaloes here as subhaloes that were at earlier times subhaloes of a different host. The fractions of these subhaloes are listed in \Tableref{tab:sublives}.
\begin{table}
\setlength\tabcolsep{2pt}
\centering\footnotesize
\caption{{\bf Subhalo population.} We list the number of subhaloes $N_{\rm S}$ used to estimate the following statistics: Backsplash Fraction $f_{\rm BS}$; fraction that swap hosts at least once $f_{\rm SH}$; the preprocessed fraction $f_{\rm PP}$; and the flyby fraction $f_{\rm FB}$. Simulations limited to those with numerous well resolved haloes across cosmic time.}
\begin{tabular}{l|ccccc}
\hline
\hline
    Name & $N_{\rm S}$ & $f_{\rm BH}$ & $f_{\rm SH}$ & $f_{\rm PP}$ & $f_{\rm FB}$ \\ 
\hline
    L40N512     & 2355  & 0.234 & 0.282 & 0.291 & 0.033 \\ 
    L210N1024   & 14774 & 0.257 & 0.193 & 0.368 & 0.028 \\ 
    L210N1536   & 46861 & 0.246 & 0.235 & 0.291 & 0.040 \\
\end{tabular}
\label{tab:sublives}
\end{table}

\par
Clearly, subhaloes are dynamic, with $\sim40-50\%$ either leaving their host halo momentarily, switching hosts entirely or just momentarily becoming a subhalo. About 1/4 of all subhaloes also leave the virial radius of their host halo before being re-accreted, with another $\sim2\%$ leaving their host entirely. Approximately $25\%$ of all long lived subhaloes truly leave one host and enter another one. The preprocessing of subhaloes, that is where a subhalo's host halo is itself accreted, is a natural outcome of hierarchical structure formation. We do not find this to be a dominant channel of accretion, though it is still significant at $30\%$. Given the resolution limits, this fraction is likely an underestimate. 

\par 
These populations may have host mass-scale dependence. For example, L40N512, which resolves lower mass host haloes and subhaloes than the L210 simulations, has a higher fraction of subhaloes that change hosts or that are flybys, though it has fewer subhaloes that have been preprocessed. We will explore the host dependence of subhalo orbits in great detail in upcoming work.

\subsubsection{How subhaloes orbit their host}\label{sec:orbits}
\begin{figure}
    \centering
    \includegraphics[width=0.49\textwidth,]{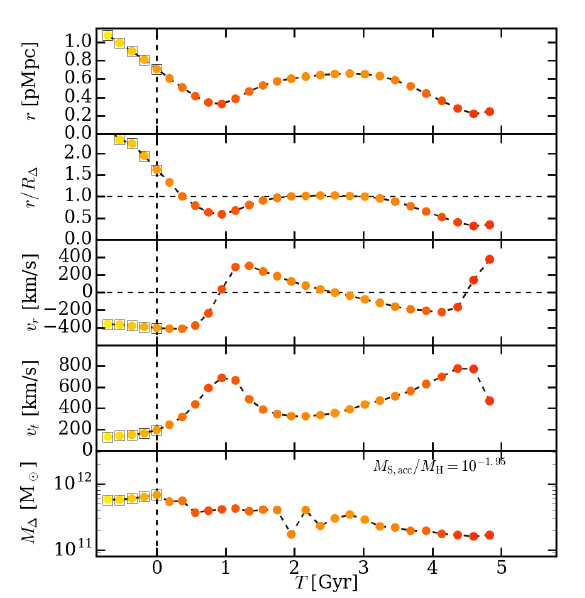}
    \caption{{\bf Sample orbit}. We show the radial distance of the subhalo from the host halo, the same radius but normalised by the virial radius of the host halo, the radial velocity, tangential velocity and mass of a subhalo that has completed a single orbit as a function of time since accretion. Here the accretion is defined as when a subhalo enters the 6D phase-space FOF envelop of the halo. The accretion mass ratio is also listed. Solid circles indicate where snapshots occur and the dashed line is the interpolated phase-space position of the subhalo. Circles highlighted by squares are points where the object is a halo, not a subhalo. Points are colour coded according to radial distance, going from yellow (distant) to red (close).}
    \label{fig:sampleorbit}
\end{figure}
\begin{figure*}
    \centering
    \includegraphics[width=0.99\textwidth,]{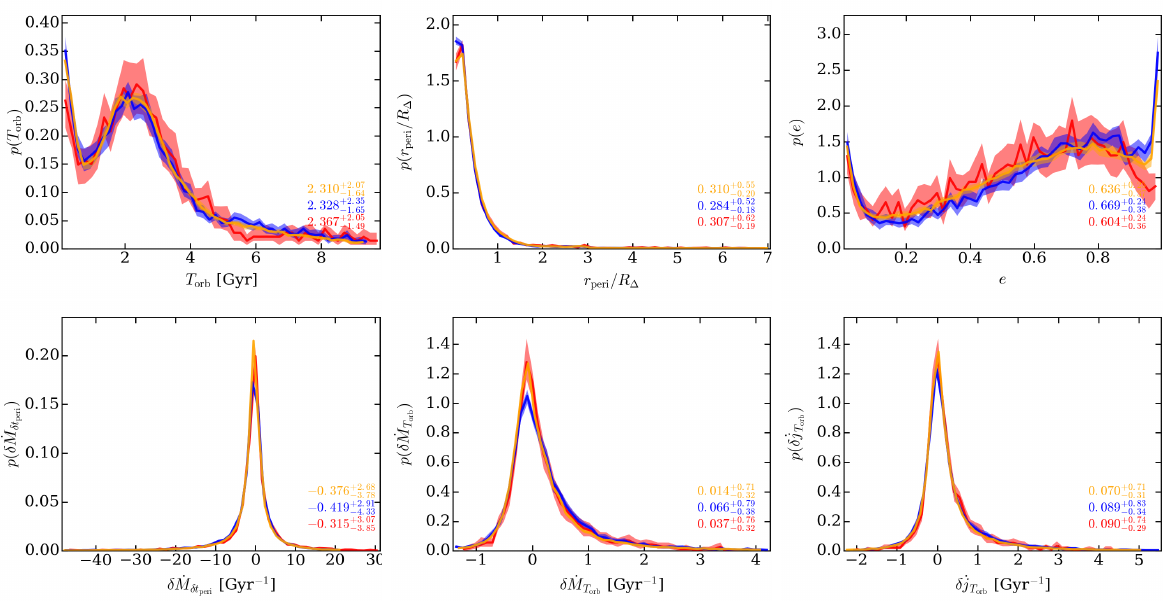}
    \caption{{\bf Orbital properties}. We show the median distribution for orbital periods, pericentric passage, and ellipticity of the orbit in the top row and the mass change over pericentric passage, mass change over an orbit and the change in orbital angular momentum over an orbit in the bottom row. We show the median normalised histograms (binned pdfs) (solid lines) along with bootstrap estimated errors (shaded regions) for each property. We also show the median value for the property along with the $16\%,84\%$ quantiles. Colour, marker and line styles are the same as in \Figref{fig:massfunc}.}
    \label{fig:orbits}
\end{figure*}
We examine the orbit of subhaloes about their host halo, interpolating their radial and tangential positions and velocities between snapshots and identify peri/apocentric passages using changes in the radial motion. An example of an orbit is shown in \Figref{fig:sampleorbit} for a subhalo that was accreted at a look back time of 5 Gyrs. Note that this subhalo is not disrupted, the curves halt $\approx5$~Gyr after accretion as this is the end of the simulation, i.e., $z=0$. 

\par 
The pericentre/apocentres are easily identifiable, giving an orbital period of 3.5 Gyr. This subhalo typically experiences gradual mass loss over its orbits, which starts principally after first pericentric passage. Although there is a minor fluctuation in the virial mass when the subhalo is near apocentre, this fluctuation is a result of the close passage of another subhalo, artificially skewing the recovered mass. There is also a kink in the mass associated from being identified by the FOF algorithm as a field halo to being identified as a subhalo by the velocity/phase-space subhalo algorithm, as FOF masses are inclusive, including substructure within the object whereas substructures have exclusive masses that do not include internal substructure. There is also a drop in mass of $10\%$ from the loss of high angular momentum material in the FOF envelop upon accretion. In general the evolution of the object is well recovered in the dense environment of a halo\footnote{Configuration-space halo finders exhibit systematic artificial decrease in assigned subhalo mass with decreasing radius \cite[][]{muldrew2011}, along with larger fluctuations than the factor of 3 change in mass seen here, with these fluctuations occurring more often.}.

\par 
The resulting probability distribution functions (pdfs) of orbital properties for subhaloes is shown in \Figref{fig:orbits}. Here we focus on the orbital period, identified between two pericentric passages, the radius of pericentre (normalised by the virial radius of the host), the orbital ellipticity $e=\tfrac{r_{\rm apo}-r_{\rm peri}}{r_{\rm apo}+r_{\rm peri}}$, and three evolutionary properties: the fractional mass loss just after pericentre passage, averaged over an orbit, and the fractional change in the orbital angular momentum change averaged over an orbit. These evolutionary properties are defined as  
\begin{align}
    \delta\dot M_{\delta t_{\rm peri}}&\equiv \frac{1}{\delta t}\frac{M(t_{\rm peri}+0.5\delta t)-M(t_{\rm peri}-0.5\delta t)}{M(t_{\rm peri}-0.5\delta t)},\\
    \delta\dot M_{T_{\rm orb}}&\equiv \frac{1}{T_{\rm orb}}\sum_{i,{\rm orb}}\left(\frac{M(t_{i+1})-M(t_{i})}{M(t_{i})}\right),\\
    \delta\dot j_{T_{\rm orb}}&\equiv \frac{1}{T_{\rm orb}}\sum_{i,{\rm orb}}\left(\frac{j(t_{i+1})-j(t_{i})}{j(t_{i})}\right).
\end{align}
Here $j\equiv |{\bf r}\times {\bf v}|$, is the specific orbital angular momentum, $\delta t$ is the time between snapshots, and the sums are for times over the orbit.

\par 
Overall, our simulations give similar pdfs for all these orbital properties. Typical orbital times are $\sim2$~Gyrs or more, so that first pericentric passage occurs roughly a Gyr after accretion. The distribution shows little variation between simulations. The few subhaloes with very short periods ($<0.2$~Gyr) result from the simple identification of peri/apocentres. We identify these points along an orbit using changes in the sign of $v_r$, the radial velocity relative to the host halo. At large radii outside the host halo where $v_r$ is small, encounters with other haloes can alter the infall velocity leading to spurious identification of peri/apocentric passage, which happens for $\approx2\%$ of all pericentres measured (these typically also $e\sim0$). Very long orbital times of $\gtrsim8$~Gyr are also a result of misidentified apocentres and are more representative of infall times.

\par 
The pericentre pdf shows typical pericentres of $\lesssim0.3R_{\Delta}$. The few subhaloes with pericentres identified at large radii $r_{\rm peri}/R_\Delta>2.0$ are due to spurious identification of pericentric passage, and occurs less than $1\%$ of the time.

\par 
The orbital ellipticity pdfs appears to be composed of two distinct populations. One at very low $e$, which drops quickly and another which is much broader that peaks at $e\sim0.7$. In general, most subhaloes are on quite elliptical orbits, with $50\%$ having $e\gtrsim0.59$. The origin of these populations is simple: orbits with $e\lesssim0.2$ arise from subhaloes that have already completed 2 pericentric passages, whereas $e\gtrsim0.2$ arise from infalling orbits, where the subhalo has completed a passage from infall to pericentre and out to apocentre. The dense environment within a halo circularises orbits. There are also a small fraction ($\lesssim3\%$) with very elliptical ($e\gtrsim0.99$) radially plunging orbits, where an apocentre has been incorrectly identified, and is actually on first infall with long orbit times. 

\par 
All the evolutionary properties show that most subhaloes do not experience large changes in either their mass or their orbit. Averaged over an orbit, a subhalo does not change mass significantly, $\delta\dot{M}_{T_{\rm orb}}\sim0$. This is not too surprising given the dynamic lives of subhaloes, where they can spent significant amounts of time outside the virial radius of their host, and where they could re-accrete some material. The figure shows that subhaloes experience mass loss at pericentric passage, with median mass loss rates of $\sim40\%$ per Gyr, though there can be significant variations in the mass loss rate. 

\par
The orbital angular momentum is generally conserved with some change in $j$ accompanying the mass loss. The positive angular momentum change can circularise an orbit. This can be seen from relating $\dot{j}_{T_{\rm orb}}$, $e$, $r_{\rm peri}$ \& $r_{\rm apo}$. If $\dot{j}_{T_{\rm orb}}>0$ then as $j=r_{\rm apo}v_{\rm apo}$, both $\dot{r_{\rm apo}}>0$ \& $\dot{v_{\rm apo}}>0$ with the same also true at pericentre. With a little bit of algebra, one can show that $\dot{j}_{T_{\rm orb}}>0$ is satisfied for 
\begin{align}
    -\frac{1-e^2}{2}\frac{\dot{r}_{\rm peri}}{r_{\rm peri}}<\dot{e}<\frac{1-e^2}{2}\frac{\dot{v}_{\rm peri}}{v_{\rm peri}}.
\end{align}
This range spans $\dot{e}<0$, corresponding to circularising orbits.

\subsubsection{Life after Accretion}
\begin{figure}
    \centering
    \includegraphics[width=0.49\textwidth,]{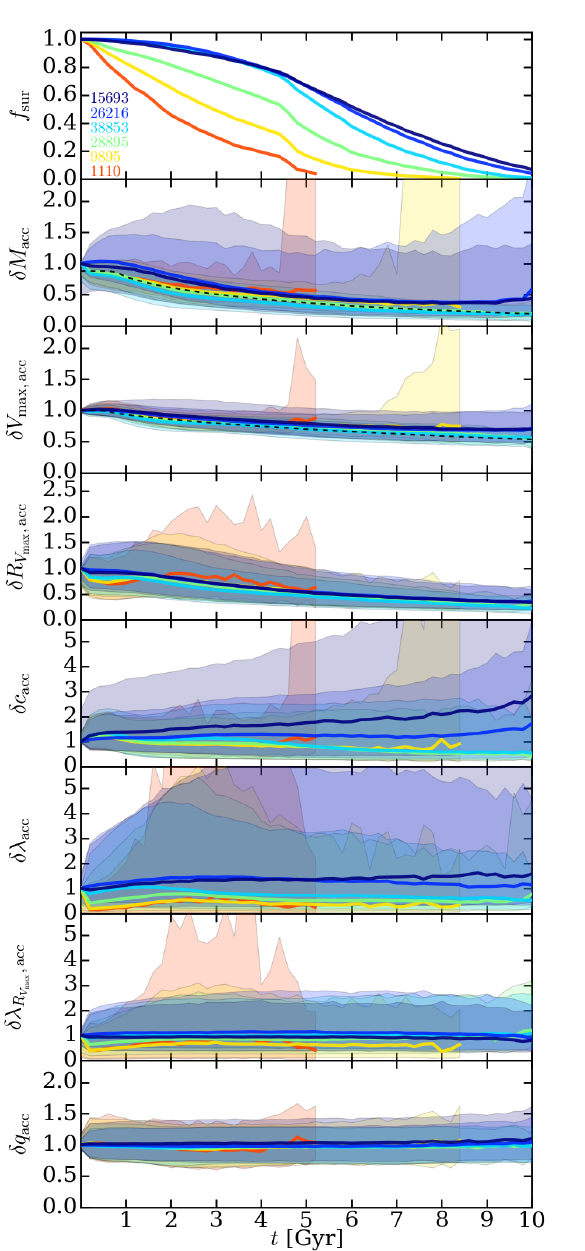}
    \caption{{\bf Subhalo evolution}. We show the average evolution of subhaloes in different accretion mass fraction bins from L210N1536 simulation. For each bin we plot the median evolution (solid thick lines) and shaded regions showing the $16\%,84\%$ quantiles. Colours go from blue to red in decreasing accretion mass ratio between the subhalo and its host. Note that the dark blue curve is for subhaloes that were larger than their host prior to accretion, and the blue curve for major/minor mergers (mass ratio of 1:1 to 1:10). We show the fraction of subhaloes remaining as a function of time from accretion $f_{\rm sur}$, the change in mass, $\vmax$, comoving $R_{\vmax}$ size, $c$, $\lambda$, $\lambda_{R_{\vmax}}$ and $q$. We also show in the top panel the number of objects in each accretion mass ratio bin. Curves end when the mass accretion ratio bin contains fewer than 20 subhaloes or till the end of the simulation at $z=0$.}
    \label{fig:avesubevo}
\end{figure}
We examine the lives of subhaloes after accretion in more detail in \Figref{fig:avesubevo}, again by following subhaloes identified at $z=0.4$. Here we only keep those objects that spend most of their life as a subhalo after accretion (removing flybys but keeping backsplash and host swapping subhaloes\footnote{A more detailed analysis of mass loss resulting from preprocessing similar to \cite{joshi2017a} is saved for future work.}). We only show results from L210N1536 as results from other simulations are similar. We calculate the fractional change of a variety of properties relative to the accretion value as a function of time, such as mass $\delta M_{\rm acc}\equiv M(t)/M_{\rm acc}$, and plot the median \& $1\sigma$ variation given by the $16\%$ and $84\%$ quantiles. We split the evolution of subhaloes based on the mass ratio between the subhalo and its host prior to accretion going from blue to red in decreasing mass ratios of $[{\color{DarkBlue}\geq1},{\color{Blue}1:10^{-1}},{\color{Cyan}10^{-1}:10^{-2}},{\color{LightGreen}10^{-2}:10^{-3}},{\color{Goldenrod}10^{-3}:10^{-4}},{\color{RedOrange}\leq10^{-4}}]$. Note that mass ratios indicative of mergers are given by dark blue curves for subhaloes which are more massive then their host prior to accretion and blue curves for major/minor mergers (mass ratio of 1:1 to 1:10). We should also point out that the subhalo sample presented here is biased towards larger mass ratios as we include all host haloes composed of $\geq5000$ particles, for which the smallest mass ratio that can be identified given the 20 particle limit is $10^{-2.4}$. Subhaloes with smaller mass fractions can only reside in the largest host haloes with $M_{\Delta}\gtrsim2\times10^{13}\Msun$.

\par
The survival fraction depends on mass ratio between the subhalo and its host. For very small mass ratios of $\lesssim10^{-3}$ (yellow and red curves) there are resolution effects at play. Subhaloes can only be identified when composed of 20 particles, thus when a subhalo drops below this threshold, the subhalo is considered to be tidally disrupted. Of greater importance is artificially shorten lifetimes due to artificial evaporation. \cite{vandenbosch2017a} found $\approx80\%$ of subhaloes in the Bolshoi+{\sc ROCKSTAR} (sub)halo catalogue are artificially disrupted, although this typically occurs for moderately resolved (sub)haloes composed of $\lesssim1000$ particles after they have lost $\gtrsim90\%$ of their accretion mass. The artificial evaporation rate is higher for less well resolved subhaloes. The mass ratio bins with $\lesssim10^{-2}$ (green to red) are likely to have artificially reduced survival fractions as these bins are dominated by subhaloes composed of $\lesssim100$ particles. 

\par
Considering these caveats, the lifetime of most objects is quite long, $\gtrsim5$~Gyr. It does appear that of the subhaloes identified at $z=0.4$, $\approx60\%$ are on long enough orbits with pericentres far enough away to avoid disruption. We will examine the orbits and the associated survival rates in an upcoming paper. 

\par
The median evolution in mass relative to the accretion $M_{\rm acc}$ indicates subhaloes initially do not significantly change mass upon accretion, with a delay of $\sim1$~Gyr, before typically losing mass at a almost linear rate. This slight delay is roughly 1/2 the typical orbit time (see \Figref{fig:orbits} in \Secref{sec:orbits}). Even major mergers (blue and dark blue curves) typically do not suffer much mass loss, gradually shrinking to $1/10^{\rm th}$ their mass over the course of Gyrs. For subhaloes with small mass ratios, accretion can remove some of the loosely bound particles assigned to the object by the FOF algorithm as seen by the small dip in mass just after accretion, particularly for poorly resolved subhaloes with artificially shallow potential wells.

\par 
The scatter in the evolution indicates how complex the lives of subhaloes can be. The upper quantile for major mergers indicate mass growth, though in such instances one could argue that neither object should be thought of as a host halo and the other as a subhalo. These mass growth instances typically occur when the subhalo/host halo tag switches between haloes and the subhalo mass reconstruction is less certain or in instances where the merger is initially a glancing one, both objects becoming field haloes for a time before re-coalescing. Even minor mergers can experience some mass growth, accreting some of the more loosely bound outer regions of a host halo. In general, the scatter is $\approx0.3$~dex save for poorly resolved subhaloes (red and yellow curves) where it can be $\gtrsim0.5$~dex, with the upper quantile occasionally indicating mass growth. Here, these instances of mass growth occur when a subhalo has momentarily left its host halo and is able to accrete mass or has merged with another subhalo. The high $84\%$ quantile lines where $\delta M_{\rm acc}>1.2$ at late times for poorly resolved haloes is principally due to subhaloes that have become haloes. These former subhaloes have longer lifetimes than similar mass subhaloes that remain with their host, dominating the median and upper quantiles at late times, resulting in the ever increasing upper quantile and upturn in the median seen. 

\par 
We fit the mass change with a simple exponential decay, 
\begin{align}
    \delta M_{\rm acc}(t)&=
    \begin{cases}
        1, t<t_o,     \\
        \exp\left[-\left(\frac{t-t_o}{\tau}\right)^\alpha\right], t\geq t_o
    \end{cases},
\end{align}
where $t_o$ is the time delay between accretion and mass loss, $\tau$ is the time scale, and $\alpha$ describes the loss rate. For mass accretion ratios of $10^{-3}<M_{\rm acc}/M_{\rm H}<10^{-1}$, we find delay times of $t_o=0.65^{+0.22}_{-0.20}$~Gyr, times scales of $\tau=4.9\pm1.0$~Gyrs and $\alpha=0.80^{+0.59}_{-0.51}$\footnote{We have neglected the $\approx10\%$ drop in mass that occurs when an object goes from being identified as a FOF field halo with an inclusive mass (that is its virial mass includes subhaloes) to being identified with the phase-space subhalo algorithm which reports an exclusive mass (that its virial mass does not include subsubstructure) along with the loss of loosely unbound high angular momentum material associated with the field halo when fitting.}. It should be noted that $\alpha$ and $\tau$ are strongly anti-correlated. The large scatter between the lower and upper quantiles is reflected in the uncertainties in the parameters. Nevertheless we find haloes with larger mass ratios have longer delay times and time scales to $1\sigma$ significance. However, given these differences are likely driven not only by accretion mass ratios but orbital parameters, we leave a more detailed analysis for later. In general, the delay times are consistent with the time scales for first pericentric approach, whereas the time scales for significant mass loss indicate that several pericentric passages are required to significantly alter the mass. 

\par
The evolution in $\vmax$ follows the mass: objects gradually become smaller over time. The evolution appears almost linear with time but to compare with the time scales of mass evolution, we fit $\delta V_{\rm max,acc}$ with the same  exponential function as mass. We find a similar though slightly smaller delay time, $t_o=0.76\pm0.11$~Gyrs and similar power-law $\alpha=0.48^{+0.93}_{-0.33}$. The key difference is the longer time-scales, $\tau=17.6\pm2.0$~Gyrs. The evolution is still not quite linear with time, $\delta V_{\rm max,acc}\approx1-x+x^2/2-x^3/6$, $x=((t-t_o)/\tau)^\alpha$. There is also less scatter, $\approx0.1$~dex, in the evolutionary paths, indicating $\vmax$ is the a better tracer of subhalo evolution than total mass. 

\par
We should note that these fits and associated time-scales are biased towards long-lived subhaloes. Due to artificially enhanced mass loss for poorly resolved subhaloes, these fits are more representative of subhaloes with mass ratios of $\gtrsim10^{-3}$. The true mass loss and evolution of smaller subhaloes requires even higher resolution simulations. A more complete follow-up of the dependence of life-times and mass loss on orbital parameters will be discussed in future work. 

\par 
As typical subhaloes lose mass and decrease in $\vmax$, well resolved subhaloes shrink as seen by the decrease in $R_{\vmax}$ in the 4th panel in \Figref{fig:avesubevo}. As subhaloes shrink, they typically become more concentrated (for well resolved haloes at least). Due to the numerical limitations, low mass ratio subhaloes have systematically enlarged $R_{\vmax}$ (see \Figref{fig:vmaxrvmax}), affecting the inferred evolution, hence it is difficult to draw physically meaningful conclusions from the average population here. The scatter in large in both $R_{\vmax}$ and $c$, approximately $50\%$. 

\par 
For mass ratios of $\lesssim10^{-2}$, subhaloes initially loose loosely bound, high angular momentum material upon accretion as indicated by the decrease in spin, both within $R_{\Delta}$ and $R_{\vmax}$. Afterwards, on average these subhaloes spin up due to tidal torques, though the scatter in $\delta \lambda_{\rm acc}$ is quite large, around $0.4$~dex, with the upper region of the upper quantiles steadily increasing with time, till the object either becomes poorly resolved or has undergone several orbits. The increase in global spin is not tracked by the spin within $R_{\vmax}$, which on average remains unchanged, in qualitative agreement with the tidal torques spinning up the outskirts of subhaloes.

\par 
The single feature that does not change is the average shape as indicated by $\delta q_{\rm acc}$. This remains around 1 with a scatter of $0.2$~dex.

\par
In general, once accreted and on bound orbits, subhaloes gradually decrease in $M$, $\vmax$, $R_{\vmax}$, become more concentrated. With each orbit, the outskirts of subhaloes are spun-up, with this high angular momentum material stripped first, while the central angular momentum changes little. 

\section{Discussion}\label{sec:discussion} 
We have given an overview of the Synthetic Universes For Surveys (SURFS) simulation suite, consisting of both pure N-Body and non-radiative hydrodynamical simulations. These simulations have large enough volumes to be useful for surveys and studying the statistical properties of haloes and high enough resolution to resolve the internal properties of dark matter haloes and subhaloes, a requirement for any investigation into cosmic structure formation and galaxy formation physics. Our high fidelity halo catalogues \& merger trees allow us to follow the orbits of subhaloes and their dynamic lives to a precision never before reached in synthetic survey simulations. Plus, our simulation volumes are large enough to study the cosmic web, and examine how the life of a halo is affected by where it resides in this web. 

\par 
We have shown numerical convergence between our various volumes and resolutions, clearly demonstrating that internal halo properties and merger histories should be limited to halo composed of $\gtrsim100$ particles. Ideally, any analysis that takes halo merger trees from simulations as input should limit it to haloes with numerically converged accretion histories. The limit for a strongly numerically converged temporal halo catalogue is even more conservative, only haloes composed of $\gtrsim500$ particles should be used. This limit affects any SAM, whereas less physically meaningful HOD models can use haloes composed of $\gtrsim20$ particles. Thus, our catalogues can be used as input to SAMs to produce numerically converged galaxy population residing in haloes down to masses of $10^{11}\Msun$ and study the orbits of galaxies in small groups up to cluster environments. Our overview highlights a selection of interesting results from haloes and subhaloes. 

\par 
An analysis of large-scale structure shows:
\begin{itemize}
    \item The cosmic web is best reconstructed with haloes of $M_\Delta\gtrsim10^{11.5}\Msun$ and requires a survey to be complete to at least $M_\Delta\geq10^{12}\Msun$. Only surveys like {\sc waves} will produce robust cosmic web reconstructions.
    \item Large haloes typically occupy knots and filaments, whereas smaller haloes reside in a larger variety of environments, spanning a broad range of distances from the nearest cosmic web filament. 
\end{itemize}

\par 
By tracing haloes at high cadence across cosmic time, we find:
\begin{itemize}
    \item Haloes smoothly grow in mass \& $\vmax$, in agreement with previous work \cite[e.g.][]{wechsler2002a,vandenbosch2002a,mcbride2009a,rodriguezpuebla2016a}. The mass history of a halo identified at $z=0$ with mass $M_{\Delta,o}$ is well characterised by $\log M(a)-=\log M_{\Delta,o}\exp\left[-(a/\beta)^\alpha\right]$, with parameters that depend weakly on the $z=0$ mass. This functional form is simpler than the recent one proposed by \cite{rodriguezpuebla2016a}. The typical cosmic evolution of a $1.6\times10^{12}\Msun$ halo has $\alpha=0.84^{+0.09}_{-0.11}$ \& $\beta=0.022^{+0.002}_{-0.003}$. The velocity scale of a halo is also well characterised by this function with $1.6\times10^{12}\Msun$ haloes having  $\alpha_{\vmax}=1.12^{+0.18}_{-0.26}$ \& $\beta_{\vmax}=0.028^{+0.07}_{-0.04}$. Mass grows faster than $\vmax$.
    \item Haloes also grow in comoving $R_{\vmax}$ till they begin to virialise, which on average occurs when these haloes are common nonlinear density peaks, i.e., when the mass variance $\sigma(M)\approx1$. Haloes continue to grow in mass as they virialise but contract in comoving $R_{\vmax}$, becoming more concentrated and spherical. This turnover point depends on mass. The exact relationship between contraction, the virial state, mass accretion rate, merger rate and the mass scale will be explored in a future paper. 
\end{itemize}

\par 
Our simulations have enough well resolved haloes for a statistical analysis of the subhalo population. Our simulations show:
\begin{itemize}
    \item Subhaloes follow power-law mass and velocity functions, $n(>f)\propto f^{-\alpha}$ with indices of $\alpha_{M}=0.83\pm0.01$ and $\alpha_{V}=2.13\pm0.03$, in agreement with previous studies \cite[e.g.][]{onions2012,rodriguezpuebla2016a}, roughly independent of redshift. The halo-to-halo scatter in the mass function is $\sigma_{\alpha_{M}}\approx0.26$. The scatter in $\vmax$ is not as well constrained given the limited dynamic range, giving an overestimated scatter of $\sigma_{\alpha_{\vmax}}\approx0.86$. The amplitude shows $0.9$~dex scatter.
    \item The number of subhaloes residing within a host halo is weakly correlated with the host halo's concentration $c$ \& spin $\lambda$. A halo with $(c,\lambda)$ $1\sigma$ (above, below) the mean values will have $60\%$ more subhaloes than similar mass halo with $c$ \& $\lambda$ $1\sigma$ below the mean. 
    \item Subhaloes are dynamic residents. Approximately $25\%$ leave their host halo momentarily, becoming a backsplash subhalo, and another $\sim25\%$ changing hosts entirely. This is in rough agreement with \cite{warnick2008a,knebe2011b,vandenbosch2017a}, with differences in the fraction of dynamic subhaloes depending on the exact definition of a halo's boundary, be it defined by critical density, mean density or some other definition. We find a moderate fraction of subhaloes, $\sim30\%$, are preprocessed, having been a subhalo of a host which itself becomes a subhalo. 
    \item Two distributions of ellipticities are observed. Subhaloes that have complete several orbits can have low $e$ values due to dynamical friction and angular momentum exchange circularising orbits. These subhaloes comprise the smaller fraction of subhaloes with only $16\%$ of the population having $e\lesssim 0.15$. In general, $e$ peaks at $\sim0.75$ and has a broad distribution. Similarly, orbital periods show two populations, short orbital times for subhaloes that have complete several orbits and a broader distribution peaked at $2.4$~Gyr. 
    \item Evolution during an orbit is on average smooth, gradually losing mass and decreasing in $\vmax$, following a delay of $\sim 0.7$~Gyr, roughly close to first pericentric passage. Most of the mass loss occurs at pericentric passage, where subhaloes can lose on average $40\%$~Gyr$^{-1}$. The time scale for exponential mass loss is $4.9\pm1.0$~Gyr, that is several pericentric passages.
    \item The central regions of long-lived subhaloes are less perturbed by tidal fields, with $\vmax$ decreasing on longer timescales of $17.6\pm2.0$~Gyr. Given the steady evolution of $\vmax$, which shows less scatter than mass, this quantity is a better tracer of subhalo evolution. 
\end{itemize}

\section{What's next?}\label{sec:conclusion} 
These results are just the tip of the proverbial iceberg: much more remains to be examined, both from the dark matter halo point of view and from the perspective of galaxy formation physics. Even here we have focused on just dark matter haloes, synthetic galaxy catalogues using SAMs will follow. Future work will explore a variety of topics. For instance, the improved subhalo tracking will necessitate updates to SAMs, particularly in dynamical friction schemes and orbital angular momentum evolution. We have hundreds of group mass haloes resolved with several hundred thousand particles containing hundreds of subhaloes and tens of low mass clusters composed of millions of particles. These objects are ideal for testing the internal structure and evolution of groups across cosmic time. 

\par
\begin{figure}
    \centering
    \includegraphics[width=0.49\textwidth,]{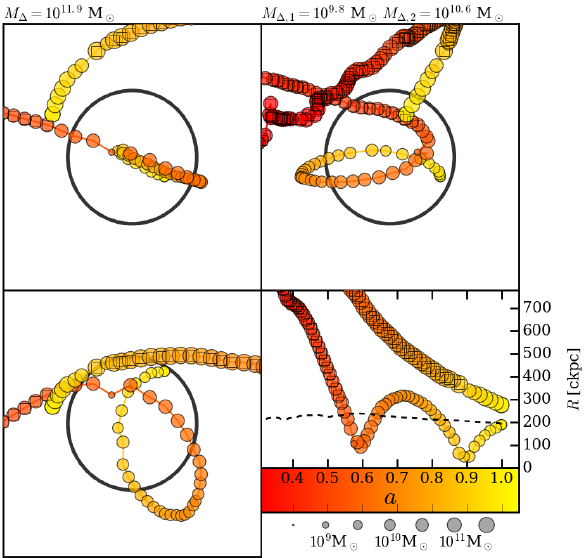}
    \caption{{\bf Orbits of Magellanic like haloes}. We show two objects orbiting/infalling onto a MW mass halo, specifically 3 projections of the orbits, $x-y$, $z-y$, \& $x-z$, and the radial position of the objects relative to the main halo as a function of $a$. Makers are colour coded by the scale factor $a$ and the marker size by the mass of the halo. In the projection plots we show the $z=0$ virial radius as a solid circle, and as a function of time in the radial plot as dashed black line. We also indicate when an object is a separate FOF halo by highlighting the point by a square.}
    \label{fig:magellanicorbit}
\end{figure}

 
We are even able to follow the evolution of Milky Way+Magellanic cloud like systems, as shown in \Figref{fig:magellanicorbit}, one of the key science goals of the {\sc waves} survey.

\par 
We have only begun identifying the fingerprints of the cosmic web on the evolution of haloes, particularly the angular momentum gas and the alignment of halo orbits to the filaments. This line of research will open up a whole new avenue for SAMs and HODs.

\par 
As work progresses, we plan to make the halo catalogues, trees and eventually semi-analytic galaxy catalogues available to the public in a searchable database and raw data, allowing the community to use {\sc surfs} data for their own research purposes. 

\section*{Acknowledgements}
We would like to thank the anonymous referee for their comments. PJE is supported by the Australia Research Council (ARC) Discovery Project Grant DP160102235. CW is supported by the Jim Buckee Fellowship. CP is supported by ARC Future Fellowship FT130100041. CL is funded by a Discovery Early Career Researcher Award DE150100618. CL also thanks the MERAC Foundation for a Postdoctoral Research Award. AR acknowledges the support of ARC Discovery Project grant DP140100395. RC is supported by the SIRF awarded by the University of Western Australia Scholarships Committee, and the Consejo Nacional de Ciencia y Tecnolog\'ia (CONACyT) scholarship No. 438594 and the MERAC Foundation. RP is supported by a University of Western Australia Scholarship. Parts of this research were conducted by the ARC Centre of Excellence for All-sky Astrophysics (CAASTRO), through project number CE110001020, and supported by the ARC Discovery Project DP160102235. This research was undertaken on Magnus at the Pawsey Supercomputing Centre in Perth, Australia and on Raijin, the NCI National Facility in Canberra, Australia, which is supported by the Australian commonwealth Government. Parts of this research were conducted by the Australian Research Council Centre of Excellence for All Sky Astrophysics in 3 Dimensions (ASTRO 3D), through project number CE170100013.

\par
The authors contributed to this paper in the following ways: PJE ran simulations and analysed the data, made the plots and wrote the bulk of the paper. CW analysed the data and wrote sections of the paper. CP ran simulations. RC \& RP assisted with with the development of software used to produce the data, {\sc VELOCIraptor} \& {\sc TreeFrog} respectively. PJE, AR, CP, and CL lead the SURFS project. All authors have read and commented on the paper.
\vspace{-10pt}
\paragraph*{Facilities} Magnus (Pawsey Supercomputing Centre), Raijin (NCI National Facility)
\vspace{-10pt}
\paragraph*{Software} Python, Matplotlib \cite[][]{matplotlib}, Scipy \cite[][]{scipy}, emcee \cite[][]{emcee}, SciKit \cite[][]{scikit}, Gadget \cite[][]{gadget2}, VELOCIraptor \cite[][Elahi et al, in prep]{elahi2011}, TreeFrog (Elahi et al, in prep)

\pdfbookmark[1]{References}{sec:ref}
\bibliographystyle{mn2e}
\bibliography{surfsintro.bbl}

\begin{thebibliography}{107}
\expandafter\ifx\csname natexlab\endcsname\relax\def\natexlab#1{#1}\fi

\bibitem[{{Allgood} {et~al}\mbox{.}(2006){Allgood}, {Flores}, {Primack},
  {Kravtsov}, {Wechsler}, {Faltenbacher}, \& {Bullock}}]{allgood2006}
{Allgood} B., {Flores} R.~A., {Primack} J.~R., {Kravtsov} A.~V., {Wechsler}
  R.~H., {Faltenbacher} A., {Bullock} J.~S., 2006, \mnras, 367, 1781

\bibitem[{{Alpaslan} {et~al}\mbox{.}(2014){Alpaslan}, {Robotham}, {Driver},
  {Norberg}, {Baldry}, {Bauer}, {Bland-Hawthorn}, {Brown}, {Cluver}, {Colless},
  {Foster}, {Hopkins}, {Van Kampen}, {Kelvin}, {Lara-Lopez}, {Liske},
  {Lopez-Sanchez}, {Loveday}, {McNaught-Roberts}, {Merson}, \&
  {Pimbblet}}]{alpaslan2014a}
{Alpaslan} M. {et~al.}, 2014, \mnras, 438, 177

\bibitem[{{Angulo} {et~al}\mbox{.}(2012){Angulo}, {Springel}, {White},
  {Jenkins}, {Baugh}, \& {Frenk}}]{angulo2012}
{Angulo} R.~E., {Springel} V., {White} S.~D.~M., {Jenkins} A., {Baugh} C.~M.,
  {Frenk} C.~S., 2012, \mnras, 426, 2046

\bibitem[{{Avila} {et~al}\mbox{.}(2014){Avila}, {Knebe}, {Pearce}, {Schneider},
  {Srisawat}, {Thomas}, {Behroozi}, {Elahi}, {Han}, {Mao}, {Onions},
  {Rodriguez-Gomez}, \& {Tweed}}]{avila2014a}
{Avila} S. {et~al.}, 2014, \mnras, 441, 3488

\bibitem[{{Baugh}(2006)}]{baugh2006}
{Baugh} C.~M., 2006, Reports on Progress in Physics, 69, 3101

\bibitem[{{Behroozi} {et~al}\mbox{.}(2015){Behroozi}, {Knebe}, {Pearce},
  {Elahi}, {Han}, {Lux}, {Mao}, {Muldrew}, {Potter}, \&
  {Srisawat}}]{behroozi2015a}
{Behroozi} P. {et~al.}, 2015, \mnras, 454, 3020

\bibitem[{{Behroozi} {et~al}\mbox{.}(2010){Behroozi}, {Conroy}, \&
  {Wechsler}}]{behroozi2010a}
{Behroozi} P.~S., {Conroy} C., {Wechsler} R.~H., 2010, \apj, 717, 379

\bibitem[{{Behroozi} {et~al}\mbox{.}(2013{\natexlab{a}}){Behroozi}, {Wechsler},
  \& {Wu}}]{rockstar}
{Behroozi} P.~S., {Wechsler} R.~H., {Wu} H.-Y., 2013{\natexlab{a}}, \apj, 762,
  109

\bibitem[{{Behroozi} {et~al}\mbox{.}(2013{\natexlab{b}}){Behroozi}, {Wechsler},
  {Wu}, {Busha}, {Klypin}, \& {Primack}}]{behroozi2013b}
{Behroozi} P.~S., {Wechsler} R.~H., {Wu} H.-Y., {Busha} M.~T., {Klypin} A.~A.,
  {Primack} J.~R., 2013{\natexlab{b}}, \apj, 763, 18

\bibitem[{{Benson}(2010)}]{benson2010b}
{Benson} A.~J., 2010, \physrep, 495, 33

\bibitem[{{Bocquet} {et~al}\mbox{.}(2016){Bocquet}, {Saro}, {Dolag}, \&
  {Mohr}}]{bocquet2016a}
{Bocquet} S., {Saro} A., {Dolag} K., {Mohr} J.~J., 2016, \mnras, 456, 2361

\bibitem[{{Bond} {et~al}\mbox{.}(1996){Bond}, {Kofman}, \&
  {Pogosyan}}]{bond1996a}
{Bond} J.~R., {Kofman} L., {Pogosyan} D., 1996, \nat, 380, 603

\bibitem[{{Boylan-Kolchin} {et~al}\mbox{.}(2009){Boylan-Kolchin}, {Springel},
  {White}, {Jenkins}, \& {Lemson}}]{boylankolchin2009}
{Boylan-Kolchin} M., {Springel} V., {White} S.~D.~M., {Jenkins} A., {Lemson}
  G., 2009, \mnras, 398, 1150

\bibitem[{{Carretero} {et~al}\mbox{.}(2015){Carretero}, {Castander},
  {Gazta{\~n}aga}, {Crocce}, \& {Fosalba}}]{carretero2015a}
{Carretero} J., {Castander} F.~J., {Gazta{\~n}aga} E., {Crocce} M., {Fosalba}
  P., 2015, \mnras, 447, 646

\bibitem[{{Cautun} {et~al}\mbox{.}(2013){Cautun}, {van de Weygaert}, \&
  {Jones}}]{cautun2013a}
{Cautun} M., {van de Weygaert} R., {Jones} B.~J.~T., 2013, \mnras, 429, 1286

\bibitem[{{Chan} {et~al}\mbox{.}(2015){Chan}, {Kere{\v s}}, {O{\~n}orbe},
  {Hopkins}, {Muratov}, {Faucher-Gigu{\`e}re}, \& {Quataert}}]{chan2015a}
{Chan} T.~K., {Kere{\v s}} D., {O{\~n}orbe} J., {Hopkins} P.~F., {Muratov}
  A.~L., {Faucher-Gigu{\`e}re} C.-A., {Quataert} E., 2015, \mnras, 454, 2981

\bibitem[{{Cole} {et~al}\mbox{.}(2000){Cole}, {Lacey}, {Baugh}, \&
  {Frenk}}]{cole2000}
{Cole} S., {Lacey} C.~G., {Baugh} C.~M., {Frenk} C.~S., 2000, \mnras, 319, 168

\bibitem[{{Colless} {et~al}\mbox{.}(2003){Colless}, {Peterson}, {Jackson},
  {Peacock}, {Cole}, {Norberg}, {Baldry}, {Baugh}, {Bland-Hawthorn}, {Bridges},
  {Cannon}, {Collins}, {Couch}, {Cross}, {Dalton}, {De Propris}, {Driver},
  {Efstathiou}, {Ellis}, {Frenk}, {Glazebrook}, {Lahav}, {Lewis}, {Lumsden},
  {Maddox}, {Madgwick}, {Sutherland}, \& {Taylor}}]{colless2003a}
{Colless} M. {et~al.}, 2003, ArXiv Astrophysics e-prints

\bibitem[{{Conroy} {et~al}\mbox{.}(2006){Conroy}, {Wechsler}, \&
  {Kravtsov}}]{conroy2006a}
{Conroy} C., {Wechsler} R.~H., {Kravtsov} A.~V., 2006, \apj, 647, 201

\bibitem[{{Contreras} {et~al}\mbox{.}(2017){Contreras}, {Padilla}, \&
  {Lagos}}]{contreras2017a}
{Contreras} S., {Padilla} N., {Lagos} C.~D.~P., 2017, ArXiv e-prints

\bibitem[{{Croton} {et~al}\mbox{.}(2016){Croton}, {Stevens}, {Tonini}, {Garel},
  {Bernyk}, {Bibiano}, {Hodkinson}, {Mutch}, {Poole}, \&
  {Shattow}}]{croton2016a}
{Croton} D.~J. {et~al.}, 2016, \apjs, 222, 22

\bibitem[{{Cui} {et~al}\mbox{.}(2012){Cui}, {Borgani}, {Dolag}, {Murante}, \&
  {Tornatore}}]{cui2012a}
{Cui} W., {Borgani} S., {Dolag} K., {Murante} G., {Tornatore} L., 2012, \mnras,
  423, 2279

\bibitem[{{Davis} {et~al}\mbox{.}(1985){Davis}, {Efstathiou}, {Frenk}, \&
  {White}}]{fof}
{Davis} M., {Efstathiou} G., {Frenk} C.~S., {White} S.~D.~M., 1985, \apj, 292,
  371

\bibitem[{{de Jong} {et~al}\mbox{.}(2014){de Jong}, {Barden}, {Bellido-Tirado},
  {Brynnel}, {Chiappini}, {Depagne}, {Haynes}, {Johl}, {Phillips}, {Schnurr},
  {Schwope}, {Walcher}, {Bauer}, {Cescutti}, {Cioni}, {Dionies}, {Enke},
  {Haynes}, {Kelz}, {Kitaura}, {Lamer}, {Minchev}, {M{\"u}ller}, {Nuza},
  {Olaya}, {Piffl}, {Popow}, {Saviauk}, {Steinmetz}, {Ural}, {Valentini},
  {Winkler}, {Wisotzki}, {Ansorge}, {Banerji}, {Gonzalez Solares}, {Irwin},
  {Kennicutt}, {King}, {McMahon}, {Koposov}, {Parry}, {Sun}, {Walton},
  {Finger}, {Iwert}, {Krumpe}, {Lizon}, {Mainieri}, {Amans}, {Bonifacio},
  {Cohen}, {Fran{\c c}ois}, {Jagourel}, {Mignot}, {Royer}, {Sartoretti},
  {Bender}, {Hess}, {Lang-Bardl}, {Muschielok}, {Schlichter}, {B{\"o}hringer},
  {Boller}, {Bongiorno}, {Brusa}, {Dwelly}, {Merloni}, {Nandra}, {Salvato},
  {Pragt}, {Navarro}, {Gerlofsma}, {Roelfsema}, {Dalton}, {Middleton}, {Tosh},
  {Boeche}, {Caffau}, {Christlieb}, {Grebel}, {Hansen}, {Koch}, {Ludwig},
  {Mandel}, {Quirrenbach}, {Sbordone}, {Seifert}, {Thimm}, {Helmi}, {trager},
  {Bensby}, {Feltzing}, {Ruchti}, {Edvardsson}, {Korn}, {Lind}, {Boland},
  {Colless}, {Frost}, {Gilbert}, {Gillingham}, {Lawrence}, {Legg}, {Saunders},
  {Sheinis}, {Driver}, {Robotham}, {Bacon}, {Caillier}, {Kosmalski}, {Laurent},
  \& {Richard}}]{dejong2014a}
{de Jong} R.~S. {et~al.}, 2014, in \procspie, Vol. 9147, Ground-based and
  Airborne Instrumentation for Astronomy V, p. 91470M

\bibitem[{{de Lapparent} {et~al}\mbox{.}(1986){de Lapparent}, {Geller}, \&
  {Huchra}}]{deLapparent1986}
{de Lapparent} V., {Geller} M.~J., {Huchra} J.~P., 1986, \apjl, 302, L1

\bibitem[{{De Lucia} \& {Blaizot}(2007)}]{delucia2007a}
{De Lucia} G., {Blaizot} J., 2007, \mnras, 375, 2

\bibitem[{{Despali} \& {Vegetti}(2016)}]{despali2016a}
{Despali} G., {Vegetti} S., 2016, ArXiv e-prints

\bibitem[{{Doroshkevich} {et~al}\mbox{.}(2004){Doroshkevich}, {Tucker},
  {Allam}, \& {Way}}]{doroshkevich2004}
{Doroshkevich} A., {Tucker} D.~L., {Allam} S., {Way} M.~J., 2004, \aap, 418, 7

\bibitem[{{Driver} {et~al}\mbox{.}(2016){Driver}, {Davies}, {Meyer}, {Power},
  {Robotham}, {Baldry}, {Liske}, \& {Norberg}}]{wavessurvey}
{Driver} S.~P., {Davies} L.~J., {Meyer} M., {Power} C., {Robotham} A.~S.~G.,
  {Baldry} I.~K., {Liske} J., {Norberg} P., 2016, The Universe of Digital Sky
  Surveys, 42, 205

\bibitem[{{Driver} {et~al}\mbox{.}(2011){Driver}, {Hill}, {Kelvin}, {Robotham},
  {Liske}, {Norberg}, {Baldry}, {Bamford}, {Hopkins}, {Loveday}, {Peacock},
  {Andrae}, {Bland-Hawthorn}, {Brough}, {Brown}, {Cameron}, {Ching}, {Colless},
  {Conselice}, {Croom}, {Cross}, {de Propris}, {Dye}, {Drinkwater}, {Ellis},
  {Graham}, {Grootes}, {Gunawardhana}, {Jones}, {van Kampen}, {Maraston},
  {Nichol}, {Parkinson}, {Phillipps}, {Pimbblet}, {Popescu}, {Prescott},
  {Roseboom}, {Sadler}, {Sansom}, {Sharp}, {Smith}, {Taylor}, {Thomas},
  {Tuffs}, {Wijesinghe}, {Dunne}, {Frenk}, {Jarvis}, {Madore}, {Meyer},
  {Seibert}, {Staveley-Smith}, {Sutherland}, \& {Warren}}]{driver2011a}
{Driver} S.~P. {et~al.}, 2011, \mnras, 413, 971

\bibitem[{{Dubinski} \& {Carlberg}(1991)}]{dubinski1991}
{Dubinski} J., {Carlberg} R.~G., 1991, \apj, 378, 496

\bibitem[{{Dubois} {et~al}\mbox{.}(2014){Dubois}, {Pichon}, {Welker}, {Le
  Borgne}, {Devriendt}, {Laigle}, {Codis}, {Pogosyan}, {Arnouts}, {Benabed},
  {Bertin}, {Blaizot}, {Bouchet}, {Cardoso}, {Colombi}, {de Lapparent},
  {Desjacques}, {Gavazzi}, {Kassin}, {Kimm}, {McCracken}, {Milliard},
  {Peirani}, {Prunet}, {Rouberol}, {Silk}, {Slyz}, {Sousbie}, {Teyssier},
  {Tresse}, {Treyer}, {Vibert}, \& {Volonteri}}]{dubois2014a}
{Dubois} Y. {et~al.}, 2014, \mnras, 444, 1453

\bibitem[{{Eardley} {et~al}\mbox{.}(2015){Eardley}, {Peacock},
  {McNaught-Roberts}, {Heymans}, {Norberg}, {Alpaslan}, {Baldry},
  {Bland-Hawthorn}, {Brough}, {Cluver}, {Driver}, {Farrow}, {Liske}, {Loveday},
  \& {Robotham}}]{eardley2015a}
{Eardley} E. {et~al.}, 2015, \mnras, 448, 3665

\bibitem[{{Elahi} {et~al}\mbox{.}(2013){Elahi}, {Han}, {Lux}, {Ascasibar},
  {Behroozi}, {Knebe}, {Muldrew}, {Onions}, \& {Pearce}}]{elahi2013a}
{Elahi} P.~J. {et~al.}, 2013, \mnras, 433, 1537

\bibitem[{{Elahi} {et~al}\mbox{.}(2011){Elahi}, {Thacker}, \&
  {Widrow}}]{elahi2011}
{Elahi} P.~J., {Thacker} R.~J., {Widrow} L.~M., 2011, \mnras, 418, 320

\bibitem[{{Foreman-Mackey} {et~al}\mbox{.}(2013){Foreman-Mackey}, {Hogg},
  {Lang}, \& {Goodman}}]{emcee}
{Foreman-Mackey} D., {Hogg} D.~W., {Lang} D., {Goodman} J., 2013, \pasp, 125,
  306

\bibitem[{{Fosalba} {et~al}\mbox{.}(2015){Fosalba}, {Crocce}, {Gazta{\~n}aga},
  \& {Castander}}]{fosalba2015a}
{Fosalba} P., {Crocce} M., {Gazta{\~n}aga} E., {Castander} F.~J., 2015, \mnras,
  448, 2987

\bibitem[{{Gao} {et~al}\mbox{.}(2012){Gao}, {Navarro}, {Frenk}, {Jenkins},
  {Springel}, \& {White}}]{gao2012}
{Gao} L., {Navarro} J.~F., {Frenk} C.~S., {Jenkins} A., {Springel} V., {White}
  S.~D.~M., 2012, \mnras, 425, 2169

\bibitem[{{Garrison-Kimmel} {et~al}\mbox{.}(2017){Garrison-Kimmel}, {Wetzel},
  {Bullock}, {Hopkins}, {Boylan-Kolchin}, {Faucher-Giguere}, {Keres},
  {Quataert}, {Sanderson}, {Graus}, \& {Kelley}}]{garrisonkimmel2017a}
{Garrison-Kimmel} S. {et~al.}, 2017, ArXiv e-prints

\bibitem[{{Gnedin} {et~al}\mbox{.}(2004){Gnedin}, {Kravtsov}, {Klypin}, \&
  {Nagai}}]{gnedin2004}
{Gnedin} O.~Y., {Kravtsov} A.~V., {Klypin} A.~A., {Nagai} D., 2004, \apj, 616,
  16

\bibitem[{{Han} {et~al}\mbox{.}(2017){Han}, {Cole}, {Frenk}, {Benitez-Llambay},
  \& {Helly}}]{han2017a}
{Han} J., {Cole} S., {Frenk} C.~S., {Benitez-Llambay} A., {Helly} J., 2017,
  ArXiv e-prints

\bibitem[{{Han} {et~al}\mbox{.}(2016){Han}, {Cole}, {Frenk}, \&
  {Jing}}]{han2016a}
{Han} J., {Cole} S., {Frenk} C.~S., {Jing} Y., 2016, \mnras, 457, 1208

\bibitem[{{Haynes} {et~al}\mbox{.}(2011){Haynes}, {Giovanelli}, {Martin},
  {Hess}, {Saintonge}, {Adams}, {Hallenbeck}, {Hoffman}, {Huang}, {Kent},
  {Koopmann}, {Papastergis}, {Stierwalt}, {Balonek}, {Craig}, {Higdon},
  {Kornreich}, {Miller}, {O'Donoghue}, {Olowin}, {Rosenberg}, {Spekkens},
  {Troischt}, \& {Wilcots}}]{haynes2011a}
{Haynes} M.~P. {et~al.}, 2011, \aj, 142, 170

\bibitem[{{Hearin} {et~al}\mbox{.}(2013){Hearin}, {Zentner}, {Berlind}, \&
  {Newman}}]{hearin2013a}
{Hearin} A.~P., {Zentner} A.~R., {Berlind} A.~A., {Newman} J.~A., 2013, \mnras,
  433, 659

\bibitem[{{Henriques} {et~al}\mbox{.}(2013){Henriques}, {White}, {Thomas},
  {Angulo}, {Guo}, {Lemson}, \& {Springel}}]{henriques2013a}
{Henriques} B.~M.~B., {White} S.~D.~M., {Thomas} P.~A., {Angulo} R.~E., {Guo}
  Q., {Lemson} G., {Springel} V., 2013, \mnras, 431, 3373

\bibitem[{{Howlett} {et~al}\mbox{.}(2017){Howlett}, {Staveley-Smith}, {Elahi},
  {Hong}, {Jarrett}, {Jones}, {Koribalski}, {Macri}, {Masters}, \&
  {Springob}}]{howlett2017b}
{Howlett} C. {et~al.}, 2017, \mnras, 471, 3135

\bibitem[{Hunter(2007)}]{matplotlib}
Hunter J.~D., 2007, Computing In Science \& Engineering, 9, 90

\bibitem[{{Johnston} {et~al}\mbox{.}(2008){Johnston}, {Taylor}, {Bailes},
  {Bartel}, {Baugh}, {Bietenholz}, {Blake}, {Braun}, {Brown}, {Chatterjee},
  {Darling}, {Deller}, {Dodson}, {Edwards}, {Ekers}, {Ellingsen}, {Feain},
  {Gaensler}, {Haverkorn}, {Hobbs}, {Hopkins}, {Jackson}, {James}, {Joncas},
  {Kaspi}, {Kilborn}, {Koribalski}, {Kothes}, {Landecker}, {Lenc}, {Lovell},
  {Macquart}, {Manchester}, {Matthews}, {McClure-Griffiths}, {Norris}, {Pen},
  {Phillips}, {Power}, {Protheroe}, {Sadler}, {Schmidt}, {Stairs},
  {Staveley-Smith}, {Stil}, {Tingay}, {Tzioumis}, {Walker}, {Wall}, \&
  {Wolleben}}]{askap}
{Johnston} S. {et~al.}, 2008, Experimental Astronomy, 22, 151

\bibitem[{Jones {et~al}\mbox{.}(2001--)Jones, Oliphant, Peterson,
  {et~al.}}]{scipy}
Jones E., Oliphant T., Peterson P., {et~al.}, 2001--, {SciPy}: Open source
  scientific tools for {Python}. [Online; accessed <today>]

\bibitem[{{Joshi} {et~al}\mbox{.}(2017){Joshi}, {Wadsley}, \&
  {Parker}}]{joshi2017a}
{Joshi} G.~D., {Wadsley} J., {Parker} L.~C., 2017, \mnras, 468, 4625

\bibitem[{{Kim} {et~al}\mbox{.}(2015){Kim}, {Park}, {L'Huillier}, \&
  {Hong}}]{kim2015a}
{Kim} J., {Park} C., {L'Huillier} B., {Hong} S.~E., 2015, Journal of Korean
  Astronomical Society, 48, 213

\bibitem[{{Klypin} {et~al}\mbox{.}(2011){Klypin}, {Trujillo-Gomez}, \&
  {Primack}}]{klypin2011a}
{Klypin} A.~A., {Trujillo-Gomez} S., {Primack} J., 2011, \apj, 740, 102

\bibitem[{{Knebe} {et~al}\mbox{.}(2011){Knebe}, {Libeskind}, {Doumler},
  {Yepes}, {Gottl{\"o}ber}, \& {Hoffman}}]{knebe2011b}
{Knebe} A., {Libeskind} N.~I., {Doumler} T., {Yepes} G., {Gottl{\"o}ber} S.,
  {Hoffman} Y., 2011, \mnras, 417, L56

\bibitem[{{Knebe} {et~al}\mbox{.}(2013){Knebe}, {Pearce}, {Lux}, {Ascasibar},
  {Behroozi}, {Casado}, {Moran}, {Diemand}, {Dolag}, {Dominguez-Tenreiro},
  {Elahi}, {Falck}, {Gottl{\"o}ber}, {Han}, {Klypin}, {Luki{\'c}},
  {Maciejewski}, {McBride}, {Merch{\'a}n}, {Muldrew}, {Neyrinck}, {Onions},
  {Planelles}, {Potter}, {Quilis}, {Rasera}, {Ricker}, {Roy}, {Ruiz},
  {Sgr{\'o}}, {Springel}, {Stadel}, {Sutter}, {Tweed}, \& {Zemp}}]{knebe2013a}
{Knebe} A. {et~al.}, 2013, \mnras, 435, 1618

\bibitem[{{Knollmann} \& {Knebe}(2009)}]{ahf}
{Knollmann} S.~R., {Knebe} A., 2009, \apjs, 182, 608

\bibitem[{{Lacey} {et~al}\mbox{.}(2016){Lacey}, {Baugh}, {Frenk}, {Benson},
  {Bower}, {Cole}, {Gonzalez-Perez}, {Helly}, {Lagos}, \&
  {Mitchell}}]{lacey2016a}
{Lacey} C.~G. {et~al.}, 2016, \mnras, 462, 3854

\bibitem[{{Lee} \& {Yi}(2013)}]{lee2013a}
{Lee} J., {Yi} S.~K., 2013, \apj, 766, 38

\bibitem[{{Liske} {et~al}\mbox{.}(2015){Liske}, {Baldry}, {Driver}, {Tuffs},
  {Alpaslan}, {Andrae}, {Brough}, {Cluver}, {Grootes}, {Gunawardhana},
  {Kelvin}, {Loveday}, {Robotham}, {Taylor}, {Bamford}, {Bland-Hawthorn},
  {Brown}, {Drinkwater}, {Hopkins}, {Meyer}, {Norberg}, {Peacock}, {Agius},
  {Andrews}, {Bauer}, {Ching}, {Colless}, {Conselice}, {Croom}, {Davies}, {De
  Propris}, {Dunne}, {Eardley}, {Ellis}, {Foster}, {Frenk}, {H{\"a}u{\ss}ler},
  {Holwerda}, {Howlett}, {Ibarra}, {Jarvis}, {Jones}, {Kafle}, {Lacey},
  {Lange}, {Lara-L{\'o}pez}, {L{\'o}pez-S{\'a}nchez}, {Maddox}, {Madore},
  {McNaught-Roberts}, {Moffett}, {Nichol}, {Owers}, {Palamara}, {Penny},
  {Phillipps}, {Pimbblet}, {Popescu}, {Prescott}, {Proctor}, {Sadler},
  {Sansom}, {Seibert}, {Sharp}, {Sutherland}, {V{\'a}zquez-Mata}, {van Kampen},
  {Wilkins}, {Williams}, \& {Wright}}]{liske2015a}
{Liske} J. {et~al.}, 2015, \mnras, 452, 2087

\bibitem[{{Madau} {et~al}\mbox{.}(2008){Madau}, {Diemand}, \&
  {Kuhlen}}]{madau2008}
{Madau} P., {Diemand} J., {Kuhlen} M., 2008, \apj, 679, 1260

\bibitem[{{Malavasi} {et~al}\mbox{.}(2017){Malavasi}, {Arnouts}, {Vibert}, {de
  la Torre}, {Moutard}, {Pichon}, {Davidzon}, {Kraljic}, {Bolzonella}, {Guzzo},
  {Garilli}, {Scodeggio}, {Granett}, {Abbas}, {Adami}, {Bottini}, {Cappi},
  {Cucciati}, {Franzetti}, {Fritz}, {Iovino}, {Krywult}, {Le Brun}, {Le
  F{\`e}vre}, {Maccagni}, {Ma{\l}ek}, {Marulli}, {Polletta}, {Pollo}, {Tasca},
  {Tojeiro}, {Vergani}, {Zanichelli}, {Bel}, {Branchini}, {Coupon}, {De Lucia},
  {Dubois}, {Hawken}, {Ilbert}, {Laigle}, {Moscardini}, {Sousbie}, {Treyer}, \&
  {Zamorani}}]{malavasi2017a}
{Malavasi} N. {et~al.}, 2017, \mnras, 465, 3817

\bibitem[{{Malavasi} {et~al}\mbox{.}(2016){Malavasi}, {Pozzetti}, {Cucciati},
  {Bardelli}, \& {Cimatti}}]{malavasi2016a}
{Malavasi} N., {Pozzetti} L., {Cucciati} O., {Bardelli} S., {Cimatti} A., 2016,
  \aap, 585, A116

\bibitem[{{McBride} {et~al}\mbox{.}(2009){McBride}, {Fakhouri}, \&
  {Ma}}]{mcbride2009a}
{McBride} J., {Fakhouri} O., {Ma} C.-P., 2009, \mnras, 398, 1858

\bibitem[{{Monaco} {et~al}\mbox{.}(2007){Monaco}, {Fontanot}, \&
  {Taffoni}}]{monaco2007a}
{Monaco} P., {Fontanot} F., {Taffoni} G., 2007, \mnras, 375, 1189

\bibitem[{{More} {et~al}\mbox{.}(2015){More}, {Diemer}, \&
  {Kravtsov}}]{more2015a}
{More} S., {Diemer} B., {Kravtsov} A.~V., 2015, \apj, 810, 36

\bibitem[{{Moster} {et~al}\mbox{.}(2010){Moster}, {Somerville}, {Maulbetsch},
  {van den Bosch}, {Macci{\`o}}, {Naab}, \& {Oser}}]{moster2010a}
{Moster} B.~P., {Somerville} R.~S., {Maulbetsch} C., {van den Bosch} F.~C.,
  {Macci{\`o}} A.~V., {Naab} T., {Oser} L., 2010, \apj, 710, 903

\bibitem[{{Muldrew} {et~al}\mbox{.}(2011){Muldrew}, {Pearce}, \&
  {Power}}]{muldrew2011}
{Muldrew} S.~I., {Pearce} F.~R., {Power} C., 2011, \mnras, 410, 2617

\bibitem[{{Murray} {et~al}\mbox{.}(2013){Murray}, {Power}, \&
  {Robotham}}]{hmfcalc}
{Murray} S.~G., {Power} C., {Robotham} A.~S.~G., 2013, Astronomy and Computing,
  3, 23

\bibitem[{{Navarro} {et~al}\mbox{.}(1997){Navarro}, {Frenk}, \& {White}}]{nfw}
{Navarro} J.~F., {Frenk} C.~S., {White} S.~D.~M., 1997, \apj, 490, 493

\bibitem[{{Navarro} {et~al}\mbox{.}(2004){Navarro}, {Hayashi}, {Power},
  {Jenkins}, {Frenk}, {White}, {Springel}, {Stadel}, \& {Quinn}}]{navarro2004}
{Navarro} J.~F. {et~al.}, 2004, \mnras, 349, 1039

\bibitem[{{Onions} {et~al}\mbox{.}(2012){Onions}, {Knebe}, {Pearce}, {Muldrew},
  {Lux}, {Knollmann}, {Ascasibar}, {Behroozi}, {Elahi}, {Han}, {Maciejewski},
  {Merch{\'a}n}, {Neyrinck}, {Ruiz}, {Sgr{\'o}}, {Springel}, \&
  {Tweed}}]{onions2012}
{Onions} J. {et~al.}, 2012, \mnras, 423, 1200

\bibitem[{{Paillas} {et~al}\mbox{.}(2016){Paillas}, {Lagos}, {Padilla},
  {Tissera}, {Helly}, \& {Schaller}}]{paillas2016a}
{Paillas} E., {Lagos} C.~D.~P., {Padilla} N., {Tissera} P., {Helly} J.,
  {Schaller} M., 2016, ArXiv e-prints

\bibitem[{Pedregosa {et~al}\mbox{.}(2011)Pedregosa, Varoquaux, Gramfort,
  Michel, Thirion, Grisel, Blondel, Prettenhofer, Weiss, Dubourg, Vanderplas,
  Passos, Cournapeau, Brucher, Perrot, \& Duchesnay}]{scikit}
Pedregosa F. {et~al.}, 2011, Journal of Machine Learning Research, 12, 2825

\bibitem[{{Peebles}(1980)}]{peebles1980}
{Peebles} P.~J.~E., 1980, {The large-scale structure of the universe}

\bibitem[{{Planck Collaboration} {et~al}\mbox{.}(2015){Planck Collaboration},
  {Ade}, {Aghanim}, {Arnaud}, {Ashdown}, {Aumont}, {Baccigalupi}, {Banday},
  {Barreiro}, {Bartlett}, \& et~al.}]{planckcosmoparams2015}
{Planck Collaboration} {et~al.}, 2015, ArXiv e-prints

\bibitem[{{Poole} {et~al}\mbox{.}(2016){Poole}, {Angel}, {Mutch}, {Power},
  {Duffy}, {Geil}, {Mesinger}, \& {Wyithe}}]{poole2016a}
{Poole} G.~B., {Angel} P.~W., {Mutch} S.~J., {Power} C., {Duffy} A.~R., {Geil}
  P.~M., {Mesinger} A., {Wyithe} S.~B., 2016, \mnras, 459, 3025

\bibitem[{{Poole} {et~al}\mbox{.}(2017){Poole}, {Mutch}, {Croton}, \&
  {Wyithe}}]{poole2017a}
{Poole} G.~B., {Mutch} S.~J., {Croton} D.~J., {Wyithe} S., 2017, ArXiv e-prints

\bibitem[{{Power} {et~al}\mbox{.}(2012){Power}, {Knebe}, \&
  {Knollmann}}]{power2012}
{Power} C., {Knebe} A., {Knollmann} S.~R., 2012, \mnras, 419, 1576

\bibitem[{{Prada} {et~al}\mbox{.}(2012){Prada}, {Klypin}, {Cuesta},
  {Betancort-Rijo}, \& {Primack}}]{prada2012a}
{Prada} F., {Klypin} A.~A., {Cuesta} A.~J., {Betancort-Rijo} J.~E., {Primack}
  J., 2012, \mnras, 423, 3018

\bibitem[{{Riebe} {et~al}\mbox{.}(2013){Riebe}, {Partl}, {Enke},
  {Forero-Romero}, {Gottl{\"o}ber}, {Klypin}, {Lemson}, {Prada}, {Primack},
  {Steinmetz}, \& {Turchaninov}}]{riebe2013}
{Riebe} K. {et~al.}, 2013, Astronomische Nachrichten, 334, 691

\bibitem[{{Rodr{\'{\i}}guez-Puebla}
  {et~al}\mbox{.}(2016){Rodr{\'{\i}}guez-Puebla}, {Behroozi}, {Primack},
  {Klypin}, {Lee}, \& {Hellinger}}]{rodriguezpuebla2016a}
{Rodr{\'{\i}}guez-Puebla} A., {Behroozi} P., {Primack} J., {Klypin} A., {Lee}
  C., {Hellinger} D., 2016, \mnras, 462, 893

\bibitem[{{Saito} {et~al}\mbox{.}(2016){Saito}, {Leauthaud}, {Hearin}, {Bundy},
  {Zentner}, {Behroozi}, {Reid}, {Sinha}, {Coupon}, {Tinker}, {White}, \&
  {Schneider}}]{saito2016a}
{Saito} S. {et~al.}, 2016, \mnras, 460, 1457

\bibitem[{{Schaye} {et~al}\mbox{.}(2015){Schaye}, {Crain}, {Bower}, {Furlong},
  {Schaller}, {Theuns}, {Dalla Vecchia}, {Frenk}, {McCarthy}, {Helly},
  {Jenkins}, {Rosas-Guevara}, {White}, {Baes}, {Booth}, {Camps}, {Navarro},
  {Qu}, {Rahmati}, {Sawala}, {Thomas}, \& {Trayford}}]{schaye2015a}
{Schaye} J. {et~al.}, 2015, \mnras, 446, 521

\bibitem[{{Scrimgeour} {et~al}\mbox{.}(2012){Scrimgeour}, {Davis}, {Blake},
  {James}, {Poole}, {Staveley-Smith}, {Brough}, {Colless}, {Contreras},
  {Couch}, {Croom}, {Croton}, {Drinkwater}, {Forster}, {Gilbank}, {Gladders},
  {Glazebrook}, {Jelliffe}, {Jurek}, {Li}, {Madore}, {Martin}, {Pimbblet},
  {Pracy}, {Sharp}, {Wisnioski}, {Woods}, {Wyder}, \& {Yee}}]{scrimgeour2012a}
{Scrimgeour} M.~I. {et~al.}, 2012, \mnras, 425, 116

\bibitem[{{Shandarin} \& {Zeldovich}(1989)}]{shandarin1989}
{Shandarin} S.~F., {Zeldovich} Y.~B., 1989, Reviews of Modern Physics, 61, 185

\bibitem[{{Sheth} {et~al}\mbox{.}(2001){Sheth}, {Mo}, \&
  {Tormen}}]{shethtormen2001}
{Sheth} R.~K., {Mo} H.~J., {Tormen} G., 2001, \mnras, 323, 1

\bibitem[{{Skibba} {et~al}\mbox{.}(2015){Skibba}, {Coil}, {Mendez}, {Blanton},
  {Bray}, {Cool}, {Eisenstein}, {Guo}, {Miyaji}, {Moustakas}, \&
  {Zhu}}]{skibba2015a}
{Skibba} R.~A. {et~al.}, 2015, \apj, 807, 152

\bibitem[{{Sousbie}(2011)}]{disperse}
{Sousbie} T., 2011, \mnras, 414, 350

\bibitem[{{Springel}(2005)}]{gadget2}
{Springel} V., 2005, \mnras, 364, 1105

\bibitem[{{Springel} {et~al}\mbox{.}(2008){Springel}, {Wang}, {Vogelsberger},
  {Ludlow}, {Jenkins}, {Helmi}, {Navarro}, {Frenk}, \& {White}}]{springel2008}
{Springel} V. {et~al.}, 2008, \mnras, 391, 1685

\bibitem[{{Springel} {et~al}\mbox{.}(2005){Springel}, {White}, {Jenkins},
  {Frenk}, {Yoshida}, {Gao}, {Navarro}, {Thacker}, {Croton}, {Helly},
  {Peacock}, {Cole}, {Thomas}, {Couchman}, {Evrard}, {Colberg}, \&
  {Pearce}}]{springel2005}
{Springel} V. {et~al.}, 2005, \nat, 435, 629

\bibitem[{{Springel} {et~al}\mbox{.}(2001){Springel}, {White}, {Tormen}, \&
  {Kauffmann}}]{subfind}
{Springel} V., {White} S.~D.~M., {Tormen} G., {Kauffmann} G., 2001, \mnras,
  328, 726

\bibitem[{{Srisawat} {et~al}\mbox{.}(2013){Srisawat}, {Knebe}, {Pearce},
  {Schneider}, {Thomas}, {Behroozi}, {Dolag}, {Elahi}, {Han}, {Helly}, {Jing},
  {Jung}, {Lee}, {Mao}, {Onions}, {Rodriguez-Gomez}, {Tweed}, \&
  {Yi}}]{srisawat2013}
{Srisawat} C. {et~al.}, 2013, \mnras, 436, 150

\bibitem[{{Stadel} {et~al}\mbox{.}(2009){Stadel}, {Potter}, {Moore}, {Diemand},
  {Madau}, {Zemp}, {Kuhlen}, \& {Quilis}}]{stadel2008}
{Stadel} J., {Potter} D., {Moore} B., {Diemand} J., {Madau} P., {Zemp} M.,
  {Kuhlen} M., {Quilis} V., 2009, \mnras, 398, L21

\bibitem[{{Staveley-Smith}(2009)}]{wallaby}
{Staveley-Smith} L., 2009, in Panoramic Radio Astronomy: Wide-field 1-2 GHz
  Research on Galaxy Evolution, p.~14

\bibitem[{{Tasitsiomi} {et~al}\mbox{.}(2004){Tasitsiomi}, {Kravtsov},
  {Gottl{\"o}ber}, \& {Klypin}}]{tasitsiomi2004a}
{Tasitsiomi} A., {Kravtsov} A.~V., {Gottl{\"o}ber} S., {Klypin} A.~A., 2004,
  \apj, 607, 125

\bibitem[{{van den Bosch}(2002)}]{vandenbosch2002a}
{van den Bosch} F.~C., 2002, \mnras, 331, 98

\bibitem[{{van den Bosch}(2017)}]{vandenbosch2017a}
{van den Bosch} F.~C., 2017, \mnras, 468, 885

\bibitem[{{van Uitert} {et~al}\mbox{.}(2016){van Uitert}, {Cacciato},
  {Hoekstra}, {Brouwer}, {Sif{\'o}n}, {Viola}, {Baldry}, {Bland-Hawthorn},
  {Brough}, {Brown}, {Choi}, {Driver}, {Erben}, {Heymans}, {Hildebrandt},
  {Joachimi}, {Kuijken}, {Liske}, {Loveday}, {McFarland}, {Miller}, {Nakajima},
  {Peacock}, {Radovich}, {Robotham}, {Schneider}, {Sikkema}, {Taylor}, \&
  {Verdoes Kleijn}}]{vanuitert2016a}
{van Uitert} E. {et~al.}, 2016, ArXiv e-prints

\bibitem[{{Vogelsberger} {et~al}\mbox{.}(2014){Vogelsberger}, {Genel},
  {Springel}, {Torrey}, {Sijacki}, {Xu}, {Snyder}, {Bird}, {Nelson}, \&
  {Hernquist}}]{vogelsberger2014a}
{Vogelsberger} M. {et~al.}, 2014, \nat, 509, 177

\bibitem[{{Wang} {et~al}\mbox{.}(2016){Wang}, {Pearce}, {Knebe}, {Schneider},
  {Srisawat}, {Tweed}, {Jung}, {Han}, {Helly}, {Onions}, {Elahi}, {Thomas},
  {Behroozi}, {Yi}, {Rodriguez-Gomez}, {Mao}, {Jing}, \& {Lin}}]{wang2016a}
{Wang} Y. {et~al.}, 2016, \mnras, 459, 1554

\bibitem[{{Warnick} {et~al}\mbox{.}(2008){Warnick}, {Knebe}, \&
  {Power}}]{warnick2008a}
{Warnick} K., {Knebe} A., {Power} C., 2008, \mnras, 385, 1859

\bibitem[{{Watson} {et~al}\mbox{.}(2013){Watson}, {Iliev}, {D'Aloisio},
  {Knebe}, {Shapiro}, \& {Yepes}}]{watson2013a}
{Watson} W.~A., {Iliev} I.~T., {D'Aloisio} A., {Knebe} A., {Shapiro} P.~R.,
  {Yepes} G., 2013, \mnras, 433, 1230

\bibitem[{{Wechsler} {et~al}\mbox{.}(2002){Wechsler}, {Bullock}, {Primack},
  {Kravtsov}, \& {Dekel}}]{wechsler2002a}
{Wechsler} R.~H., {Bullock} J.~S., {Primack} J.~R., {Kravtsov} A.~V., {Dekel}
  A., 2002, \apj, 568, 52

\bibitem[{Welker(2015)}]{welkerthesis}
Welker C., 2015, Theses, {Universit{\'e} Pierre et Marie Curie - Paris VI}

\bibitem[{{Zel'Dovich}(1970)}]{zeldovich1970}
{Zel'Dovich} Y.~B., 1970, \aap, 5, 84

\bibitem[{{Zheng} {et~al}\mbox{.}(2005){Zheng}, {Berlind}, {Weinberg},
  {Benson}, {Baugh}, {Cole}, {Dav{\'e}}, {Frenk}, {Katz}, \&
  {Lacey}}]{zheng2005a}
{Zheng} Z. {et~al.}, 2005, \apj, 633, 791

\bibitem[{{Zolotov} {et~al}\mbox{.}(2012){Zolotov}, {Brooks}, {Willman},
  {Governato}, {Pontzen}, {Christensen}, {Dekel}, {Quinn}, {Shen}, \&
  {Wadsley}}]{zolotov2012a}
{Zolotov} A. {et~al.}, 2012, \apj, 761, 71

\end{thebibliography}
\appendix

\section{Mass function}\label{sec:halomassfunc}
\begin{figure}
    \centering
    \includegraphics[width=0.49\textwidth,]{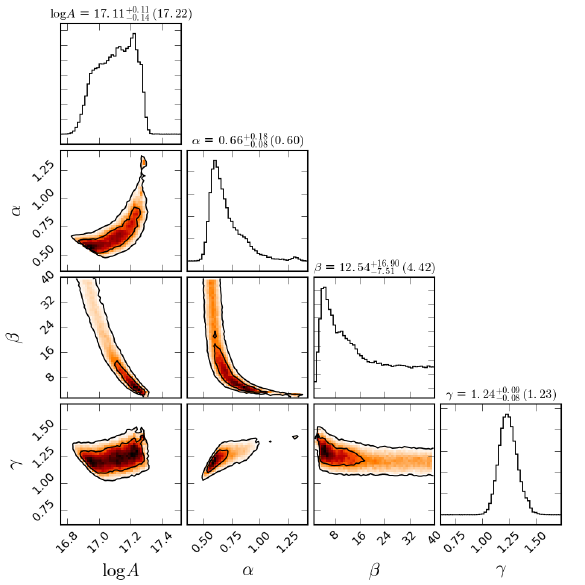}
    \caption{{\bf HMF parameter distribution}. The parameter distribution from fitting the $z=0$ HMF of the L210N1536 simulation. For each quantity we plot the median upper and lower quantiles and the value with the maximum likelihood in parenthesis.}
    \label{fig:massfit}
\end{figure}
Following numerous studies \cite[e.g.][]{watson2013a,poole2016a}, we fit the halo mass function 
\begin{align}
    \frac{dn}{d\ln M_\Delta}&=\frac{\rho_{\rm crit}}{M_\Delta} f(\sigma,z)\left(\frac{d\ln\sigma^{-1}}{d\ln M_\Delta}\right),
\end{align}
where $f(\sigma,z)$ is the scaled mass function, which here we assume a universal (i.e., redshift independent) functional form of 
\begin{align}
    f(\sigma,z)=A\left(\beta/\sigma+1\right)^\alpha e^{-\gamma/\sigma^2}.
\end{align}
Here $A,\beta$ are normalisation parameters and $\alpha,\gamma$ are shape parameters. We determine the pdfs of these model parameters using {\sc emcee} with the log likelihood given by
\begin{equation}
    \ln L= -1/2 \sum_i \left(\frac{n_i - n_{{\rm model},i}}{\sigma_i}\right)^2-\sum_i\ln{\sqrt{2\pi}\sigma_i},
\end{equation}
where $i$ is the sum over the bins in the binned differential mass function, $\sigma_{i}=\sqrt{n_i+1/4}+1/2$ is the associated modified Poisson error \cite[][]{watson2013a} and  
\begin{equation}
    n_{\rm model_{i}}=\int \frac{dn}{d\ln M_\Delta} d\ln M_{\Delta},
\end{equation}
the integral of the differential mass function over the bin.

\par
Typically, studies fit the binned mass function derived from the entire simulation volume, correcting for missing power on scales larger than the simulation volume. Though this might give the most precise fit to simulation data, it is not an accurate representation of the true HMF\footnote{Additionally, some studies fit the FOF mass, others the mass enclosing densities of $\rho=\Delta\rho_{\rm crit}$, or  $\rho=\Delta\rho_{\rm m}$, with $\Delta=200$ to $\Delta(z)$ a function of redshift based on spherical collapse.}. Baryon physics can significantly alter dark matter haloes. Moreover, the resulting pdfs of the parameters will be missing the scatter coming from cosmic variance (and missing large-scale power). Given these issues we do not attempt to account for missing large-scale power nor cosmic variance. We would argue that fitting the HMF from an pure N-Body simulation to within $1\%$ is not a meaningful task unless compared to other simulations in a careful check of softening lengths, hydrodynamical parameters, and subgrid physics \cite[e.g.][are just a few of the studies that show the significant impact baryons have on dark matter, from dark matter concentration to dark matter mass]{gnedin2004,zolotov2012a,cui2012a,chan2015a,bocquet2016a,despali2016a,garrisonkimmel2017a}. 

\par 
Instead we simply fit the binned differential mass function sampled at the 100 different mass scales, the first 50 largest haloes to capture the exponential turn-over, and the next equally spaced in $\log M$. The results of the fitting procedure from the {\sc emcee} package \cite[][]{emcee} using uninformed priors are shown in \Figref{fig:massfit}. Our parameter pdfs are qualitatively the same as any other work, there are degeneracies between normalisation parameters, $\log A$ and $\beta$, between the normalisation parameters and the shape parameter $\alpha$. The exponential shape parameter $\gamma$ is weakly correlated with other parameters. The normalisation parameter $\beta$, has a long tail due to the number of haloes constraining the turn-over from power-law to exponential, and the most likely value of $\beta=4.42$. Our results are broadly in agreement with \cite{watson2013a}, as well as the fits from \cite{poole2016a}, which were made at much higher redshifts.

\par
We note that the exploration of parameters and their degeneracies is also done for other fits, such as the subhalo mass function or the halo growth, using {\sc emcee}. 

\section{Cosmic Growth \& Resolution} \label{sec:cosmicgrowthapp}
\begin{figure}
    \centering
    \includegraphics[width=0.49\textwidth,]{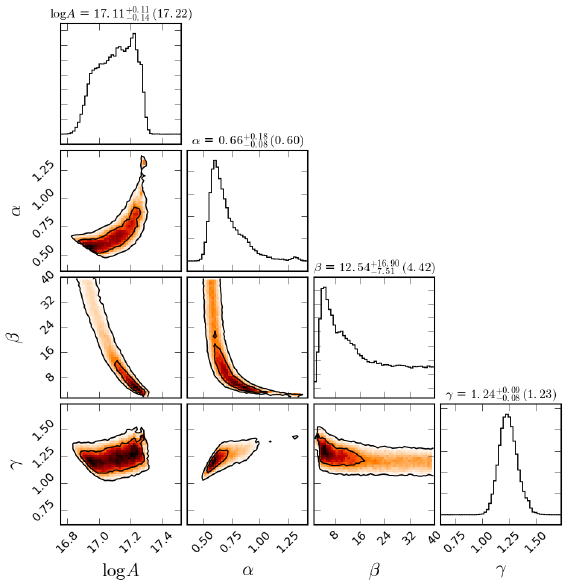}
    \caption{{\bf Halo Evolution of L40N512}. The average halo evolution similar to \Figref{fig:avehaloevo} but for L40N512.}
    \label{fig:avehaloevoL40N512}
\end{figure}

\begin{figure}
    \centering
    \includegraphics[width=0.49\textwidth,]{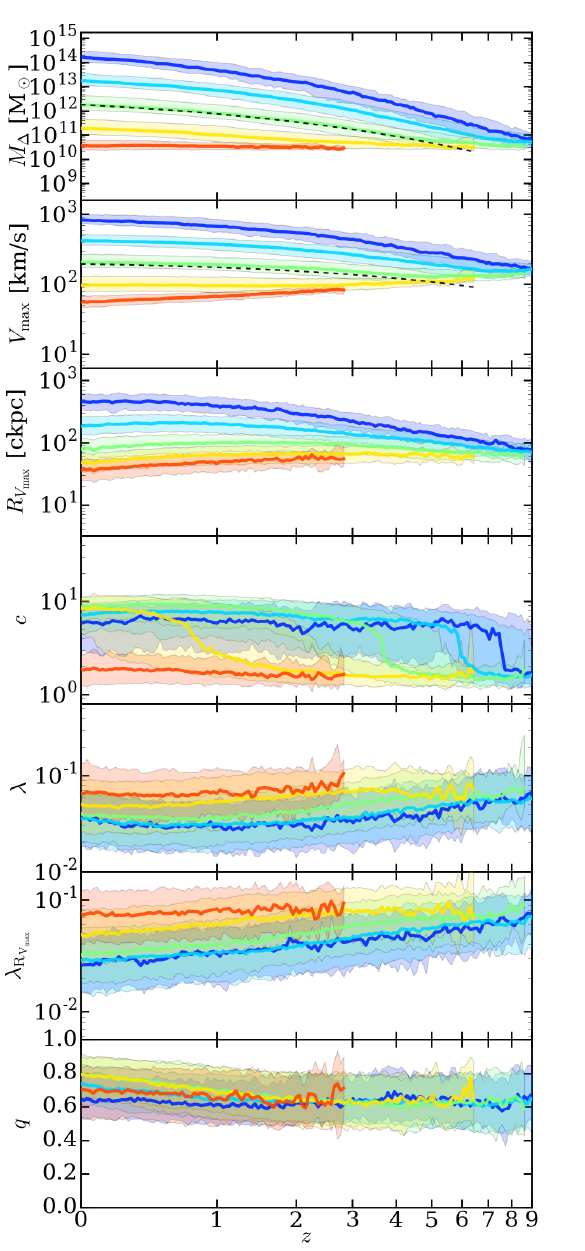}
    \caption{{\bf Halo Evolution of L210N1024}. The average halo evolution similar to \Figref{fig:avehaloevo} but for L210N1024.}
    \label{fig:avehaloevoL210N1024}
\end{figure}

We show the growth of cosmic structure of different simulations here, specifically our higher mass resolution, smaller volume simulation (\Figref{fig:avehaloevoL40N512}) and our lower resolution L210N1024 simulation (\Figref{fig:avehaloevoL210N1024}). The median growth of haloes observed in L210N1536 is reproduced by L40N512 for well resolved haloes in both simulations, that is mass bins of $\gtrsim10^{12}\Msun$. The halo mass bin of $10^{10}-10^{11}\Msun$ (red curve), which contains haloes composed of $\sim162-1619$ particles in L40N512, reproduces the evolutionary behaviour of large haloes, which is not seen in L210N1536 due to resolution effects. 

\par 
A similar impact of reducing our resolution is seen in L210N1024. Here the $10^{11}-10^{12}\Msun$ (yellow curve) shows little evolution beyond a redshift of 2, and the $10^{10}-10^{11}\Msun$ (red curve) shows no evolution. In contrast L210N1536 $10^{11}-10^{12}\Msun$ haloes begin to evolve from $z\sim4$ onwards. 

\end{document}